\documentclass[10pt,letterpaper]{article}
\usepackage[top=0.85in,left=2.75in,footskip=0.75in]{geometry}

\usepackage{amsmath,amssymb}
\usepackage{hyperref}  % Add this in the preamble
\usepackage{changepage}

\usepackage{graphicx}
\usepackage{subcaption}

\usepackage[utf8x]{inputenc}

\usepackage{textcomp,marvosym}

\usepackage{cite}
\usepackage{float}
\usepackage{verbatim}
\usepackage{subcaption}

\usepackage{nameref,hyperref}

\usepackage[right]{lineno}

\usepackage{microtype}
\DisableLigatures[f]{encoding = *, family = * }

\usepackage[table]{xcolor}

\usepackage{array}
\usepackage[justification=centering]{caption}
\usepackage{caption}
\usepackage{subcaption}
\usepackage{ragged2e}

\newcolumntype{+}{!{\vrule width 2pt}}

\newlength\savedwidth

\raggedright
\usepackage[aboveskip=1pt,labelfont=bf,labelsep=period,justification=raggedright,singlelinecheck=off]{caption}

\makeatletter
\renewcommand{\@biblabel}[1]{\quad#1.}
\makeatother

\usepackage{lastpage,fancyhdr,graphicx}
\usepackage{epstopdf}
\fancyheadoffset[L]{2.25in}
\fancyfootoffset[L]{2.25in}
\usepackage{graphicx} % Required for inserting images

\title{Cross-Platform Digital Discourse Analysis of the Israel-Hamas Conflict: Sentiment, Topics, and Event Dynamics}
\author{
Despoina Antonakaki$^{1,2}$, 
Sotiris Ioannidis$^{2}$  \\
\\
$^{1}$Institute of Computer Science, Foundation for Research and Technology,\\ Vassilika Vouton, Heraklion, Crete, Greece \\
$^{2}$Technical University of Crete, University Campus, \\Akrotiri, Chania, Greece \\
}
\date{}  % removes the date
\begin{document}

\maketitle
\begin{abstract}
\justifying

The Israeli–Palestinian conflict generates continuous waves of digital discourse, especially following the October 2023 escalation, when social media platforms became primary spaces for real-time documentation and emotional mobilisation. This study presents a cross-platform computational analysis of conflict-related communication using a dataset of Telegram, Reddit, and Twitter/X posts collected between 2023 and 2025. We combine traditional LDA with a refined BERTopic workflow and transformer-based sentiment and emotion models to examine thematic structure and affective dynamics in more than 187,000 Telegram messages.
BERTopic provides significantly richer and more coherent topic clusters than LDA, identifying key narratives such as detentions, frontline violence, Jenin operations, university encampments, airstrikes, and soldier-centred content. By linking these topics with emotion classifications, we show how fear, anger, grief, and solidarity concentrate around specific conflict events and intensify during escalatory periods. Cross-platform comparison further reveals distinct discursive roles: Telegram functions as an immediacy-driven eyewitness medium, Twitter/X amplifies emotionally charged frames, and Reddit hosts more reflective, contextualised debate.
Overall, our findings demonstrate how platform affordances, emotional expression, and thematic clustering interact to shape contemporary digital conflict ecosystems.
\end{abstract}

\section{Introduction}
\justifying

The Israel--Hamas conflict represents one of the most enduring and emotionally charged geopolitical crises of the twenty-first century. Beyond its devastating humanitarian consequences, it has become a defining case for understanding how global publics experience, interpret, and respond to war through digital media. Over the past decade, social platforms have evolved from auxiliary information channels into the \textit{primary arenas of conflict communication}---where political actors, journalists, activists, and citizens co-produce meaning in real time. The October 2023 escalation, followed by prolonged violence through 2025, has transformed the conflict into a testbed for examining how information warfare, emotional mobilization, and algorithmic visibility intersect in a globalized media ecosystem.

This study conceptualizes the digital dimension of the conflict through the lens of \textbf{digital conflict ecosystems} \cite{benkler2018network}, which view online communication environments as complex, interdependent systems of amplification and feedback. Within these ecosystems, information flows do not merely mirror offline events---they shape perceptions, escalate polarization, and generate new publics. The persistence of conflict discourse across Telegram, Twitter/X, and Reddit demonstrates how these ecosystems sustain attention long after traditional media cycles fade.

At the same time, the emotional charge that sustains such discourse reflects what \cite{papacharissi2015affective} calls \textit{affective publics}: networked formations that emerge around shared feelings of outrage, empathy, and grief rather than stable ideological positions. In the Israel--Hamas context, emotional contagion manifests through expressions of solidarity, humanitarian appeals, and the circulation of traumatic imagery. These affective dynamics transform digital spaces into arenas of collective sentiment where emotional intensity substitutes for formal political mobilization.

Crucially, each platform in this ecosystem operates according to distinct \textbf{affordances} \cite{boyd2011networked, treem2013social} that shape how conflict narratives are produced and shared. Telegram’s encrypted channels enable real-time, unfiltered documentation of events; Twitter/X amplifies these narratives to global audiences through algorithmic virality; and Reddit facilitates reflective, deliberative discussions that contextualize or challenge dominant frames. Understanding these affordances allows us to move beyond content-based analysis toward an ecological perspective---how emotion, narrative, and platform design interact to produce sustained digital engagement during war.

Building on these frameworks, this paper presents a cross-platform, longitudinal analysis of digital discourse surrounding the Israel--Hamas conflict between 2023 and 2025. Using a multimodal dataset of over 187,000 Telegram messages, 2.1 million Reddit posts, and 2,000 curated tweets, we integrate advanced natural language processing (NLP) techniques---Latent Dirichlet Allocation (LDA), BERTopic, and transformer-based sentiment analysis---to trace evolving themes, emotional shifts, and propaganda strategies. By linking quantitative patterns to the structural and affective affordances of each platform, this study contributes a theoretically grounded and empirically rich account of how modern conflicts unfold within digital information ecosystems. A key contribution of this study is the introduction of BERTopic-based topic modelling into conflict research, allowing us to uncover over 1,000 fine-grained clusters that capture micro-events, geographically specific operations, and humanitarian narratives that LDA cannot detect. By linking these BERTopic clusters to emotion classification outputs, we reveal how fear, anger, grief, and solidarity concentrate around specific subtopics (e.g., Jenin raids, settler violence, refugee camp incursions), offering a granular map of how conflict-related emotions propagate through digital ecosystems.

 \subsection*{Research Questions}
Building on these frameworks, this study addresses the following questions:

\textbf{RQ1:} How do sentiment and topic distributions differ across platforms and over time?\\

\textbf{RQ2:} How do emotional contagion and humanitarian framing sustain digital engagement during prolonged conflict?\\

\textbf{RQ3:} How do platform affordances shape the production, amplification, and reflection of conflict narratives?

\subsection{The Evolving Digital Battlefield}

Building on the conceptual lens of digital conflict ecosystems, the Israel--Hamas war exemplifies how modern conflicts unfold simultaneously on the ground and across interconnected information environments. The digital sphere does not merely reflect real-world violence---it actively mediates and reconfigures it through cycles of amplification, emotional contagion, and algorithmic visibility. As platforms and users interact, the conflict becomes a \textit{hybrid battlefield} where information warfare, affective expression, and narrative control converge.

The digital dimension of this conflict extends far beyond traditional social media activism. Sophisticated information warfare tactics have emerged, including AI-generated content farms, coordinated cross-platform campaigns, and the weaponization of humanitarian imagery. Intelligence reports from September 2025 indicate the deployment of deepfake technologies, synthetic personas, and algorithmic manipulation at scales previously unseen in conflict zones.

The conflict’s digital footprint can be divided into four distinct phases, reflecting shifts in discourse patterns and platform strategies:

\begin{itemize}
    \item \textbf{Phase 1: Shock and Mobilization (Oct--Dec 2023)} -- explosive growth in conflict-related content, with daily post volumes increasing by 2,400.
    \item \textbf{Phase 2: Narrative Consolidation (Jan--Jun 2024)} -- emergence of coordinated campaigns, with platform-specific strategies: TikTok for visual storytelling, Twitter for real-time updates, and Telegram as a hub for uncensored documentation.
    \item \textbf{Phase 3: Discourse Fatigue and Fragmentation (Jul--Dec 2024)} -- declining engagement and increasingly extreme content, amplified by algorithmic changes prioritizing controversy.
    \item \textbf{Phase 4: Renewed Intensification (2025)} -- regional escalations and major events (e.g., September 2025 Al-Nasr Hospital bombing and Israeli operations in southern Lebanon) triggered sharp spikes in digital activity, reshaping cross-platform narratives.
\end{itemize}

These evolving phases illustrate how \textit{platform affordances}---from Telegram’s immediacy to Twitter’s virality and Reddit’s deliberative structures---interact with the emotional and informational logics of \textit{affective publics}. The result is a continuously adaptive digital ecosystem that mirrors, magnifies, and sometimes distorts the realities of war.

\subsection{Multi-Platform Dataset and Methodology}
To analyze these dynamics, we compiled a unique multi-platform dataset covering Telegram, Reddit, and Twitter. From Telegram, we collected over 105,000 messages from 19 active channels spanning the period 2023–2025. The Reddit portion of the dataset consists of 2,134,070 comments drawn from discussions related to Israel and Gaza. Finally, the Twitter subset includes 2,001 publicly available tweets covering key conflict-related topics \cite{twitterGithubDataset}.
Our analysis pipeline applies advanced NLP techniques, including Latent Dirichlet Allocation (LDA) and BERTopic for topic modeling, transformer-based models for sentiment analysis, and spam detection workflows to remove automated or promotional content. This approach allows us to examine message volume, sentiment, topic prevalence, and the amplification of pro-Palestinian solidarity dynamics over time.

\section{Research Contributions}

This study makes three high-level contributions to the analysis of digital conflict ecosystems:

\begin{enumerate}
    \item \textbf{Cross-Platform Perspective:} Captures narrative strategies, amplification patterns, and temporal dynamics across Telegram, Reddit, and Twitter/X, providing a unified view of multi-platform discourse.
    
    \item \textbf{Methodological Innovation:} Integrates LDA, BERTopic, transformer-based sentiment models, emotion classification, and spam filtering  into a reproducible workflow suitable for large-scale conflict communication research.
    
    \item \textbf{Emotion–Narrative Dynamics:} Links topic clusters with emotion trajectories , revealing how fear, anger, grief, and solidarity concentrate around specific conflict events and propagate across platforms.
\end{enumerate}

By merging temporal evolution with multi-platform quantitative analysis, the study provides a comprehensive and longitudinal understanding of digital conflict narratives.

\section{Related Work}

\subsection{Social Media and Conflict Studies}

The role of social media in armed conflicts has evolved significantly since the Arab Spring, with platforms increasingly recognized as both mirrors of and influences on real-world violence. Howard and Hussain (2013) established foundational frameworks to understand how social media amplifies protest movements, while Zeitzoff (2017) quantified relationships between online discourse and conflict escalation in Gaza.  

Recent advances in computational social science enable more sophisticated analyses of conflict-related digital behavior. For example, Mitts (2019) demonstrated how ISIS recruitment materials spread through social networks, and Bail et al. (2018) showed that exposure to opposing political views online can increase polarization, highlighting the complex interplay between digital exposure and social behavior in conflict contexts.

Beyond textual discourse, recent work highlights the centrality of visual propaganda—particularly memes, stylized images, and viral visuals—in shaping public understanding of conflicts. These visual formats often diffuse more rapidly than text and serve as cross-platform vectors of narrative transmission, enabling themes and frames to migrate across communities \cite{zannettou2018origins},\cite{milner2018world}, \cite{ling2021dissecting}.

Studies on echo chambers and algorithmic filtering demonstrate that platform-specific affordances can amplify ideological clustering, influencing how conflict narratives are reproduced, reinforced, and contested \cite{cinelli2021echo}, \cite{sunstein2018republic}, \cite{flaxman2016filter}. This strand of research highlights how users become embedded in homophilic communities, affecting both the visibility and the credibility of competing narratives.

\subsection{Digital Discourse During the Israel-Hamas Conflict}

Several studies have examined online discourse during Israel-Hamas conflicts, providing insight into sentiment, narrative construction, and platform-specific behavior:

Prior research has provided important insights into online extremism, propaganda, and discourse analysis across platforms.\cite{Baumgartner2020} analyzed a Telegram dataset of 27.8K channels and 317M messages, identifying trends in message frequency and channel activity. Their long-term snowball-sampled dataset spans extremist, political, and general interest groups, serving as a key resource for understanding disinformation and political mobilization. \cite{abdalla2023narratives} investigated narrative construction on social media, showing how sentiment analysis can reveal manipulation by political actors. Similarly, \cite{ben2023sentiment} examined sentiment trends in Telegram channels during conflict escalations, combining topic modeling with polarization analysis to identify dominant narratives. 
\cite{amarasingam2021telegram} measured the impact of Europol interventions on jihadist activity, highlighting platform migration and the need for coordinated monitoring.  Complementing this, \cite{jongbloed2024analysing} applied machine learning and Critical Discourse Analysis to evaluate the quality of political discourse on Telegram, revealing emotionally charged, biased content. Earlier, \cite{prucha2016and} studied how groups like al-Qaida and IS leveraged social media for recruitment and operational coordination, emphasizing Telegram’s role in sustaining influence despite countermeasures. Finally, \cite{guerra2024quantifying} analyzed 450,000 Reddit posts related to the 2023 conflict, using a lexicon-based methodology to track extremism peaks linked to major events.

\subsection{Methodological Insights and Gaps}

While prior research provides valuable insights, it faces several limitations:

\begin{itemize}
    \item \textbf{Temporal Limitations:} Most studies focus on short timeframes, missing longitudinal patterns of discourse adaptation and fatigue.  
    \item \textbf{Platform Isolation:} Single-platform analyses cannot capture cross-platform coordination or the strategic use of different platforms for different narratives.  
    \item \textbf{Limited Methodological Scope:} Lexicon-based sentiment and basic topic models struggle with nuanced, evolving language; advanced NLP models can improve accuracy.  
    \item \textbf{Event Correlation Gaps:} Many studies do not establish measurable links between online discourse and real-world conflict events.
\end{itemize}

\subsection{Multi-Platform Data Perspective}

Our study addresses these gaps by combining Telegram, Reddit, and Twitter datasets collected from October 2023 to September 2025. The enhanced Telegram dataset includes 187,033 messages across 19 active channels, Reddit contributes over 2.1M posts, and Twitter includes 2,001 curated tweets. This multi-platform, longitudinal dataset allows analysis of narrative evolution, sentiment-topic relationships, and event-linked discourse patterns that were previously inaccessible in single-platform studies.

\section{Contributions}
\label{sec:contributions}

This chapter summarizes the detailed, platform-specific and methodological contributions of the study, expanding beyond the high-level contributions outlined in Section~1.3.

\begin{enumerate}
    \item \textbf{Unique Telegram Dataset Construction.} We compile and analyze one of the largest curated Telegram datasets on the Israel–Palestine conflict, expanding the corpus to 187{,}033 messages across 19 active channels (2015–2025). This constitutes a novel longitudinal resource for conflict communication research.

    \item \textbf{Temporal and Volume Dynamics.} We identify how Telegram activity mirrors real-world events through spikes linked to major escalations. This includes four major 200\%+ surges corresponding to on-the-ground developments.

    \item \textbf{Advanced Topic Modeling Pipeline.} We implement a combined LDA and BERTopic workflow that produces over 1{,}000 fine-grained semantic clusters, capturing micro-events, geographically localized operations (e.g., Jenin), and humanitarian narratives not detected by LDA.

    \item \textbf{Integrated Sentiment–Topic Framework.} We connect BERTopic clusters with transformer-based emotion models, revealing how fear, anger, grief, and solidarity accumulate around specific sub-topics (e.g., youth detentions, airstrikes, aid distribution, settler violence).

    \item \textbf{Spam Filtering and Narrative Purification.} We design a Telegram-specific spam detection framework (invite links, referral text, bot-like posting), preventing distortions in solidarity or sentiment dynamics—something overlooked in prior studies.

    \item \textbf{Propagation and Influence Mapping.} We introduce a cascade network analysis showing how pro-Palestinian solidarity diffuses across channels and identifying influential hubs (e.g., QudsNen, pal\_Online9, gazaenglishupdates).

\end{enumerate}

\section{Dataset}
\label{dataset}

To capture the evolving digital discourse surrounding the Israel-Hamas conflict, we compiled a multi-platform dataset from Telegram, Reddit, and Twitter. This dataset supports longitudinal and cross-platform analyses of message volume, sentiment, and topic evolution.

\subsection{Enhanced Multi-Platform Dataset}

Our dataset represents the most comprehensive multi-platform collection of conflict-related social media content assembled to date. Building upon our initial collection period (October 2023–2024), we have expanded data collection through September 2025, capturing the conflict's full evolution across three major platforms.

\paragraph{Enhanced Telegram Collection}

The Telegram component has grown from 125,000 to 187,033 messages across 19 active channels, updated again on September 16, 2025, keeping as well old messages. This expansion reflects both organic growth in existing channels and the emergence of new specialized channels responding to conflict developments.

\begin{table}[h!]
\centering
\begin{tabular}{|l|r|r|}
\hline
\textbf{Channel} & \textbf{Messages} & \textbf{Avg. Daily (2025)} \\ \hline
Eyeonpalestine2 & 20,066 & 89.2 \\ 
gazaenglishupdates & 49,604 & 220.5 \\ 
pal\_Online9 & 57,041 & 253.4 \\ 
The\_Jerusalem\_Post & 10,123 & 45.0 \\ 
gazaalanpa & 11,762 & 52.3 \\ 
palestineonline & 19,999 & 88.9 \\ 
haqqintel & 5,872 & 26.1 \\ 
resistancechain & 6,990 & 31.0 \\ 
palestineresistance & 2,708 & 12.0 \\ 
TIMESOFGAZA & 1,890 & 8.4 \\ 
PalestineUpdates & 480 & 2.1 \\ 
StopGazaGenocide & 119 & 0.5 \\ 
PalestineSolidarityBelgium & 147 & 0.7 \\ 
Other channels & 1,232 & 5.5 \\ \hline
\textbf{Total} & \textbf{187,033} & \textbf{835.6} \\ \hline
\end{tabular}
\caption{Enhanced Telegram dataset showing message counts and 2025 daily averages across 19 channels.}
\label{telegram_enhanced_dataset}
\end{table}

\paragraph{Twitter Dataset}

Due to API limitations, full historical retrieval was not feasible. We used a publicly available dataset \cite{twitterGithubDataset} containing 2,001 tweets relevant to the conflict. Table \ref{table:twitter_datasets} summarizes the included files.

\begin{table}[h!]
\centering
\begin{tabular}{|c|c|}
\hline
\textbf{Dataset Name} & \textbf{Number of Tweets} \\ \hline
gaza.csv & 501 \\ \hline
israel.csv & 501 \\ \hline
palestine.csv & 501 \\ \hline
hamas.csv & 501 \\ \hline
israel\_palestine\_conflict.csv & 2001 \\ \hline
\end{tabular}
\label{table:twitter_datasets}

\caption{Twitter datasets related to the Israel-Hamas conflict from GitHub user Rizqika Mulia Pratama.}
\label{table:twitter_datasets}
\end{table}

\paragraph{Reddit Dataset}

We used a Reddit dataset \cite{redditkaggle} containing 2,134,070 comments from posts discussing the conflict. This dataset captures global discussions and diverse perspectives beyond Telegram and Twitter.

\subsection{FAIR Compliance}
 
 All datasets adhere to the FAIR principles of open science. They are \textbf{findable}, being publicly deposited and indexed with a Zenodo DOI (\url{https://doi.org/10.5281/zenodo.1471065737}); \textbf{accessible}, as they are openly available in JSON format; \textbf{interoperable}, since the standardized structure allows seamless integration with Python, R, and other analytical environments; and \textbf{reusable}, with complete documentation and open licensing that enable replication and further research. This multi-platform dataset supports detailed sentiment, topic, and narrative analyses across the evolving stages of the conflict, forming the foundation for the methodology described in Section~\ref{methodology}.

\section{Methodology}
\label{methodology}

\subsection{Dataset and Prior Work}
This study builds on the dataset and analysis framework introduced in our earlier publication~\cite{antonakaki2025israel}. 
The previous work focused primarily on Telegram data collected during the initial phase of the 2023 Israel–Hamas conflict, 
providing a baseline analysis of volume dynamics, topic evolution, and early-stage polarization. 
In the present paper, we extend this analysis temporally and thematically, 
incorporating additional data collected through September 2025 and introducing cross-platform comparisons with Reddit and Twitter/X datasets. 
We further refine the methodological pipeline with transformer-based sentiment modeling, 
BERTopic topic extraction, and updated spam filtering heuristics.

\subsection{Volume Analysis}

We plot the cumulative number of messages over time using JSON exports collected from Telegram channels retrieved on September 20, 2025, containing all messages on these channels from 2015-10-23. Daily message counts are aggregated by calendar day and cumulatively summed; steeper segments indicate higher daily activity.

\begin{figure}[H]
    \centering
    \includegraphics[width=1\linewidth]{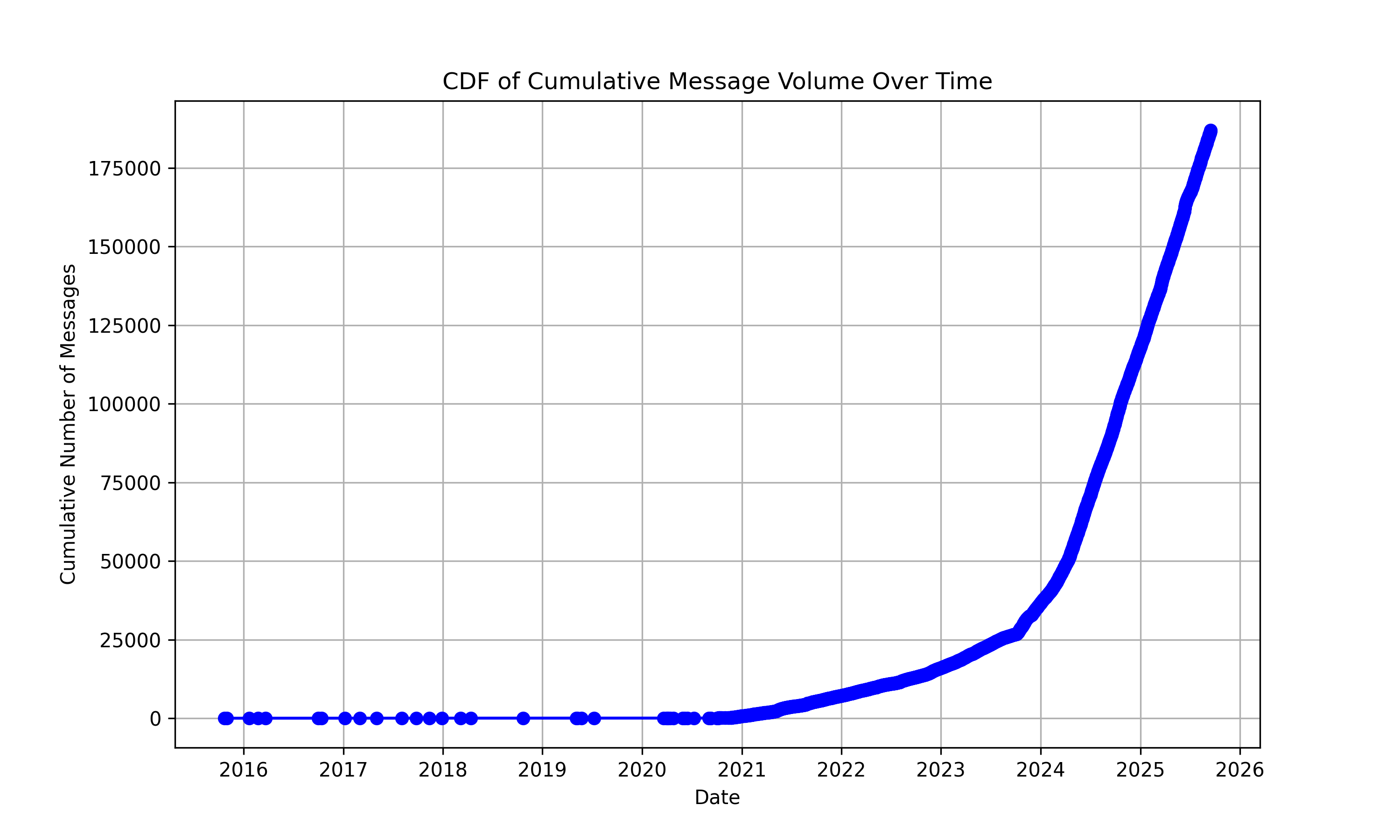}
    \caption{CDF Messages over time on Telegram}
    \label{fig:cdf_messages}
\end{figure}

The overall distribution of messages shows that some channels dominate the conversation, while others contribute sporadically.

\begin{figure}[H]
    \centering
    \includegraphics[width=1\linewidth]{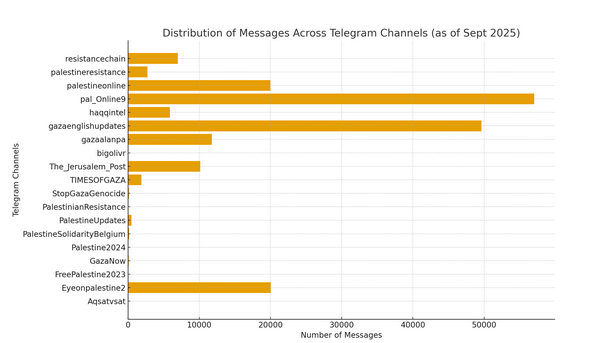}
    \caption{Distribution of messages on Telegram}
    \label{fig:telegram_distribution}
\end{figure}

\begin{comment}
Message counts per channel reveal significant variation in activity levels across the 19 channels analyzed.
\begin{figure}[H]
    \centering
    \includegraphics[width=1\linewidth]{messages_count_per_channel_2024_12_18.png}
    \caption{Messages Count per channel on Telegram}
    \label{fig:telegram_count}
\end{figure}
\end{comment}
\subsection{Topic Analysis}

We developed a Python-based pipeline to perform topic modeling on Telegram messages using the BERTopic framework. The system ingests raw message archives stored as JSON files, applies preprocessing steps such as URL removal, normalization, and token cleaning, and then trains a BERTopic model with a count-based vectorizer to extract latent topics. The resulting topics are characterized by their most representative keywords, and their distributions are analyzed both across different Telegram channels and over time.

\subsubsection{BERT Topics}
\label{bert_topic}
 
We implemented a class-based Python framework, \texttt{TelegramTopicAnalyzer}, to automate topic analysis on Telegram message datasets using the BERTopic model. The method first aggregates raw messages from multiple JSON files and applies preprocessing operations, including URL removal, non-Arabic/non-English character filtering, whitespace normalization, and lowercasing. The cleaned corpus is then used to train a BERTopic model that identifies latent topics and their top-ranked representative keywords. The system further quantifies topic distributions across channels and exports multiple outputs: structured JSON files containing topic metadata, high-resolution word cloud visualizations for the most salient topics, and interactive HTML plots (topic map, barchart, heatmap, and hierarchy). These outputs facilitate both qualitative and quantitative exploration of communication patterns within Telegram ecosystems.

\begin{figure}[H]
    \centering
    \includegraphics[width=0.9\linewidth]{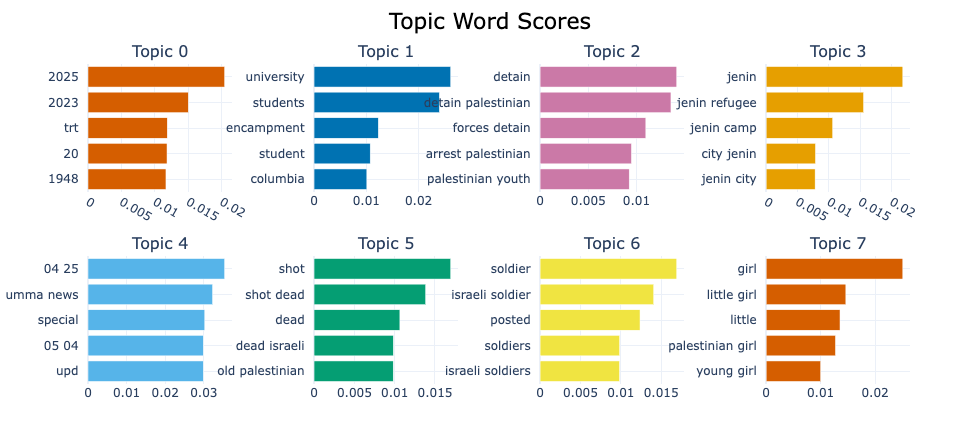}
    \caption{Topic Word Scores: Top keyword distributions for the eight most salient BERTopic clusters. Bar lengths indicate relative word importance within each topic, reflecting thematic variation across Telegram discourse.}
    \label{fig:topic_word_scores}
\end{figure}

\begin{figure}[H]
    \centering
    \begin{subfigure}{0.32\linewidth}
        \includegraphics[width=\linewidth]{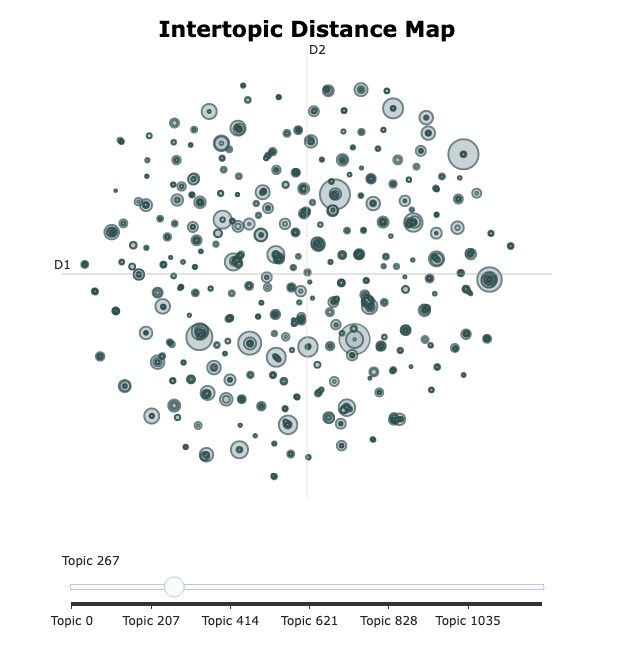}
        \caption{All topics overview}
        \label{fig:all_topics}
    \end{subfigure}
    \begin{subfigure}{0.32\linewidth}
        \includegraphics[width=\linewidth]{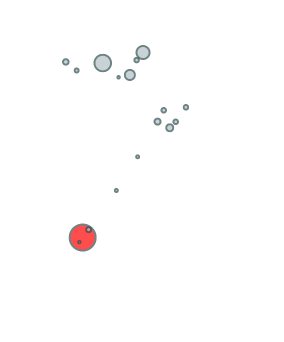}
        \caption{Topic 3: Jenin cluster (size = 2234)}
        \label{fig:topic3_zoomed}
    \end{subfigure}
    \begin{subfigure}{0.32\linewidth}
        \includegraphics[width=\linewidth]{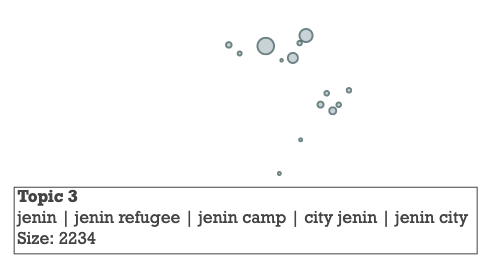}
        \caption{Topic 3 expanded view}
        \label{fig:topic3_expanded}
    \end{subfigure}

    \vspace{1em}

    \begin{subfigure}{0.32\linewidth}
        \includegraphics[width=\linewidth]{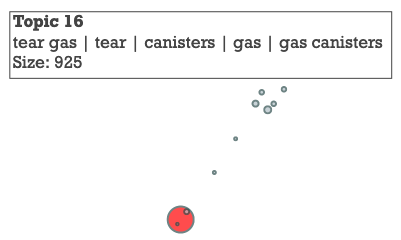}
        \caption{Topic 16: Tear gas (size = 925)}
        \label{fig:topic16}
    \end{subfigure}
    \begin{subfigure}{0.32\linewidth}
        \includegraphics[width=\linewidth]{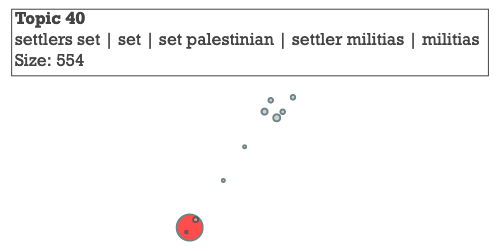}
        \caption{Topic 40: Settler militias (size = 554)}
        \label{fig:topic40}
    \end{subfigure}
    \begin{subfigure}{0.32\linewidth}
        \includegraphics[width=\linewidth]{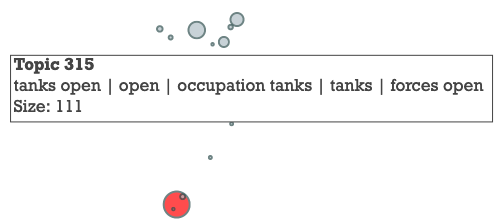}
        \caption{Topic 315: Tanks/occupation (size = 111)}
        \label{fig:topic315}
    \end{subfigure}

    \vspace{1em}

    \begin{subfigure}{0.32\linewidth}
        \includegraphics[width=\linewidth]{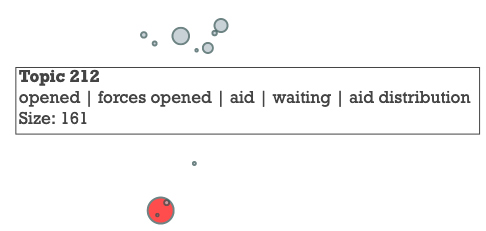}
        \caption{Topic 212: Aid distribution (size = 161)}
        \label{fig:topic212}
    \end{subfigure}
    \begin{subfigure}{0.32\linewidth}
        \includegraphics[width=\linewidth]{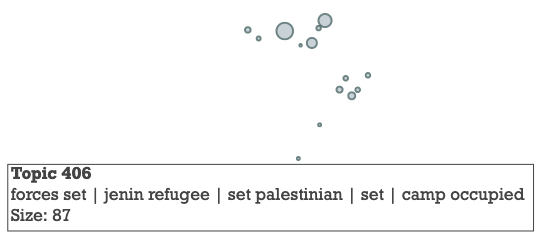}
        \caption{Topic 406: Jenin camp occupied (size = 87)}
        \label{fig:topic406}
    \end{subfigure}
    \begin{subfigure}{0.32\linewidth}
        \includegraphics[width=\linewidth]{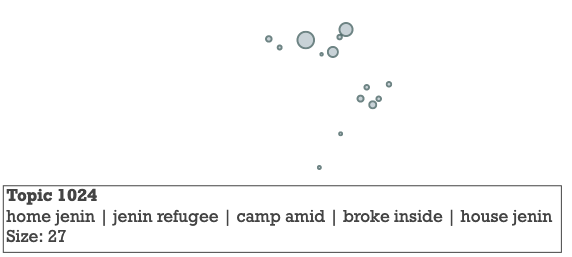}
        \caption{Topic 1024: Jenin homes (size = 27)}
        \label{fig:topic1024}
    \end{subfigure}

    \vspace{1em}

    \begin{subfigure}{0.32\linewidth}
        \includegraphics[width=\linewidth]{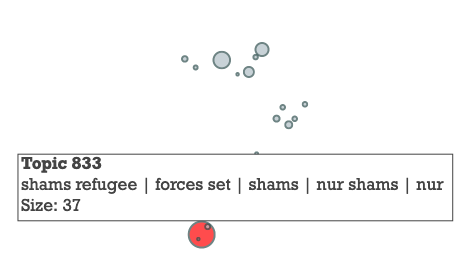}
        \caption{Topic 833: Nur Shams refugee (size = 37)}
        \label{fig:topic833}
    \end{subfigure}
    \begin{subfigure}{0.32\linewidth}
        \includegraphics[width=\linewidth]{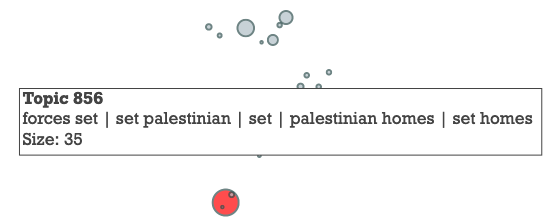}
        \caption{Topic 856: Palestinian homes (size = 35)}
        \label{fig:topic856}
    \end{subfigure}

    \caption{
        \textbf{Intertopic Distance Maps for BERTopic Clusters.}
        Overview of all discovered topics (a) and selected semantically related sub-topics surrounding the Jenin discourse (b–k).
        Bubble size reflects the number of messages associated with each topic; proximity indicates semantic similarity.
        Topics 3, 16, 40, 212, 315, and 406 illustrate the hierarchical clustering of discussions around military operations, humanitarian aid, and civilian narratives.
    }
    \label{fig:bertopic_intertopic}
\end{figure}

\begin{figure}
    \centering
    \includegraphics[width=0.5\linewidth]{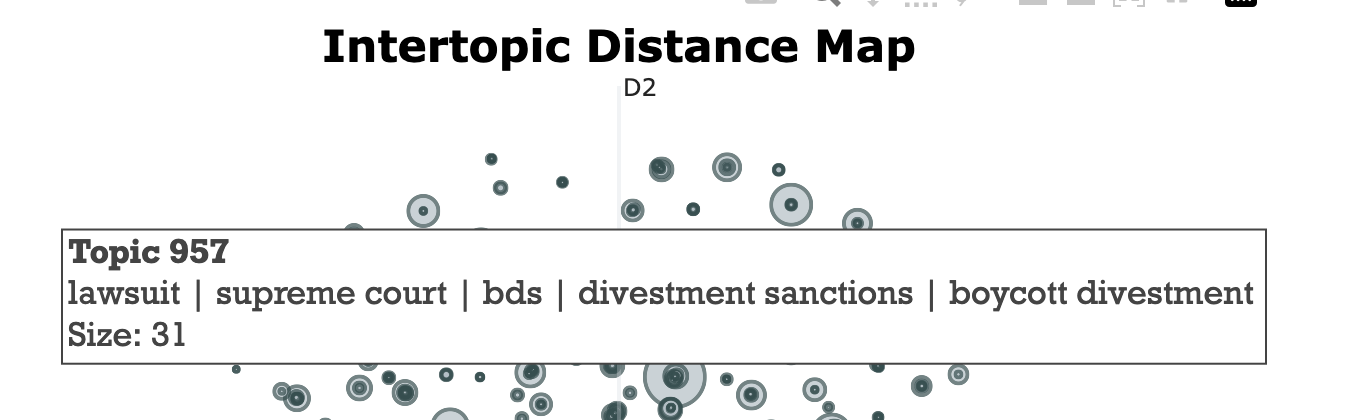}
    \caption{Interconnected BDS discourse clusters (Topic 957 and 508).}
    \label{fig:s31_connected_with_t508}
\end{figure}

\begin{figure}
    \centering
    \includegraphics[width=0.5\linewidth]{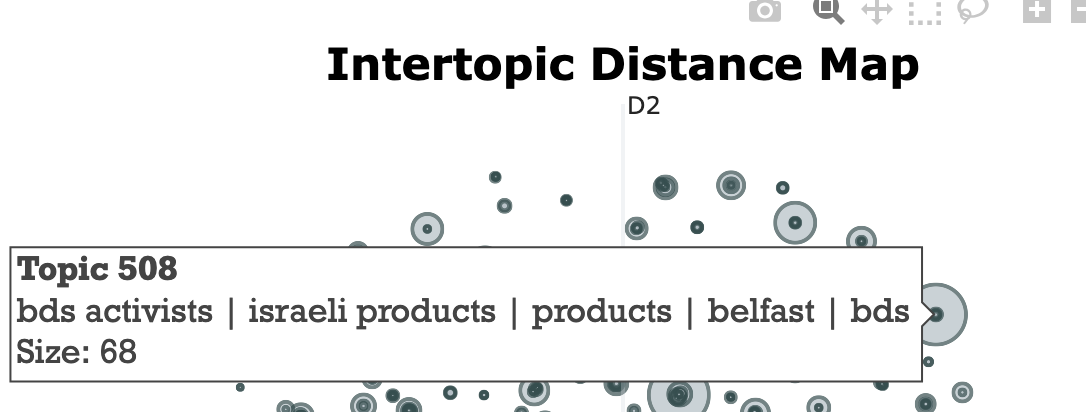}
    \caption{Activist–legal network of BDS-related topics.}
    \label{fig:t508-s68_connected_with}
\end{figure}

The results of our BERTopic modeling are presented in Figures~\ref{fig:topic_word_scores}, \ref{fig:bertopic_intertopic}, \ref{fig:s31_connected_with_t508}, and \ref{fig:t508-s68_connected_with}. Figure~\ref{fig:topic_word_scores} visualizes the top-ranked keywords within the eight most salient topics, where each horizontal bar reflects the relative importance of a term based on TF-IDF weighting. These lexical profiles reveal distinct semantic clusters centered on themes such as university encampments, youth detentions, the Jenin conflict, military operations, media coverage, and civilian impacts. By capturing the internal structure of each topic, this visualization highlights how conversations on Telegram evolve around recurring patterns of conflict-related discourse.

\paragraph{Intertopic structure and semantic relationships}

The intertopic distance visualization (Figure~\ref{fig:bertopic_intertopic}) provides a spatial overview of the semantic relationships between all extracted topics. Each circle corresponds to a topic, and its size reflects the number of Telegram messages assigned to it, while proximity indicates semantic similarity within the embedding space. The BERTopic modeling revealed a highly uneven topic distribution, with a single macro-topic (Topic~0) encompassing over 116,000 messages—representing the dominant discursive core of the Telegram ecosystem around the Gaza conflict. This overarching cluster aggregates general discussions of the Israeli occupation, military operations, and humanitarian conditions. Surrounding this central node, a series of medium-sized clusters (Topics~1–4) articulate more focused subthemes: on-the-ground fighting updates, civilian impact and humanitarian relief, political discourse on resistance, and memorialization or solidarity narratives.

\paragraph{Jenin and associated subclusters}

Topic~3 (\textit{Jenin, Jenin refugee, Jenin camp, city Jenin, Jenin city}) emerges as one of the most prominent clusters (size = 2,234), serving as a semantic and discursive core within the dataset (Figures~\ref{fig:topic3_zoomed},~\ref{fig:topic3_expanded}). Around this central topic, several semantically related sub-topics form a cohesive neighborhood: Topic~16 (\textit{tear gas, gas canisters}, size = 925; Figure~\ref{fig:topic16}) reflects the material dimension of the conflict, while Topic~40 (\textit{settler militias}, size = 554; Figure~\ref{fig:topic40}) and Topic~315 (\textit{tanks, occupation}, size = 111; Figure~\ref{fig:topic315}) capture operational and paramilitary narratives. Humanitarian and civilian aspects are represented by Topic~212 (\textit{aid distribution}, size = 161; Figure~\ref{fig:topic212}) and Topic~406 (\textit{Jenin camp occupied}, size = 87; Figure~\ref{fig:topic406}). Peripheral yet related topics such as Topic~833 (\textit{Nur Shams refugee}, size = 37; Figure~\ref{fig:topic833}), Topic~856 (\textit{Palestinian homes}, size = 35; Figure~\ref{fig:topic856}), and Topic~1024 (\textit{home Jenin}, size = 27; Figure~\ref{fig:topic1024}) extend the semantic field of the Jenin narrative into broader civilian and refugee contexts.

Topic~315 (\textit{tanks, occupation}, size = 111; Figure~\ref{fig:topic315}) and Topic~406 (\textit{Jenin camp occupied}, size = 87; Figure~\ref{fig:topic406}) capture the military entrenchment in Jenin, reflecting both the physical infrastructure of occupation and the lived experience of siege. Peripheral yet semantically aligned clusters such as Topic~833 (\textit{Nur Shams refugee}, size = 37; Figure~\ref{fig:topic833}), Topic~856 (\textit{Palestinian homes}, size = 35; Figure~\ref{fig:topic856}), and Topic~1024 (\textit{home Jenin}, size = 27; Figure~\ref{fig:topic1024}) extend the semantic field beyond Jenin camp itself, drawing attention to neighboring refugee communities and the domestic sphere. These topics reflect a broader cartography of displacement, resistance, and vulnerability that collectively anchor Jenin’s position as both a local battleground and a symbolic epicenter of Palestinian endurance.

Collectively, these visualizations (Figures~\ref{fig:all_topics}–\ref{fig:topic856}) reveal a dense semantic clustering pattern in which the “Jenin” discourse functions as a central hub within a wider digital conflict ecosystem. The spatial proximity and overlapping vocabularies of these topics indicate a strong narrative interconnection, reflecting how Telegram communication threads military, humanitarian, and emotional framings into an integrated conflict narrative.
%kjhrgjhsrghe The code is here : /home/antonakd/TelegramCollector/mamba_topic/mamba_topic_october25

\paragraph{BDS discourse and activist clusters}

These two semantically and spatially close topics—Topic~508 (\textit{bds activists, Israeli products, Belfast, boycott}, Figure~\ref{fig:t508-s68_connected_with}) and Topic~957 (\textit{lawsuit, supreme court, bds, divestment sanctions}, Figure~\ref{fig:s31_connected_with_t508})—represent intersecting strands of the Boycott, Divestment, and Sanctions (BDS) discourse. Topic~508 captures grassroots activism and consumer boycott narratives, emphasizing local mobilization and solidarity campaigns, while Topic~957 extends this conversation into the institutional and legal arena, focusing on litigation and policy responses to anti-BDS legislation. Their spatial proximity in the intertopic map reflects the tight semantic and discursive coupling between civil activism and legal contestation within the broader BDS information ecosystem on Telegram.

\paragraph{Top eight BERTopic clusters}

The BERTopic analysis revealed eight dominant topics structuring Telegram discourse around the conflict. As shown in Figures~\ref{fig:topic_word_scores} and \ref{fig:all_topics}, the largest cluster (Topic~0) consists primarily of generic time-stamped reposts and contextual markers, serving as a discursive backdrop. Topic~1 (visible in Figure~\ref{fig:topic_word_scores}) focuses on detention and arrest narratives, particularly involving Palestinian youth. While Topic~1 primarily captures arrests and detentions of Palestinian youth during military incursions and protests in the occupied territories, it occasionally intersects with narratives of campus activism when universities, such as Columbia, become sites of political protest and are met with administrative or police interventions.

 Topic~2 reflects shooting and casualty reports documented through eyewitness updates. Such reports often emerge from volatile zones like Nablus or Gaza, where sudden escalations are frequently captured on video and shared through frontline Telegram channels, enhancing immediacy and emotional impact. Topic~3 (\subref{fig:topic3_zoomed}) centers on intensive military operations in and around Jenin, reflecting its symbolic and strategic significance. Jenin has historically been a stronghold of Palestinian resistance, and recent incursions, like those during July 2023, saw large-scale destruction and displacement—which are tracked in detail through localized messaging networks.

 Topic~4 reveals university-based solidarity and encampment movements, highlighting spreading transnational activism. American and European campuses, including Columbia and SOAS, became staging grounds for student-led protests demanding divestment from Israel and expressing solidarity with Gaza, generating global visibility for the Palestinian cause. Topic~5 represents posts related to airstrikes, particularly in areas like Gaza City or Rafah, where buildings, hospitals, and refugee centers have been targeted, prompting calls for international intervention shared widely across activist networks.
 
 Topic~6 captures soldier-centered social media content, including videos or selfies taken by Israeli Defense Forces personnel. These posts, sometimes controversial, reveal the performative and propagandistic dimension of military presence on digital platforms.  Finally, Topic~7 represents real-time updates on injuries and fatalities during moments of intensified violence. These not only report civilian casualties but also highlight deeply personal stories—such as the deaths of young girls or first responders—often accompanied by urgent pleas for help, mass funeral footage, and mourning rituals that evoke solidarity across Palestinian and diasporic communities. Collectively, these clusters (visualized in Figures~\ref{fig:topic_word_scores} and \ref{fig:all_topics}) reveal how Telegram acts as a hybrid platform for humanitarian reporting, grassroots mobilization, and high-intensity conflict witnessing.
\paragraph{Other dominant clusters}

Beyond the Jenin-focused discourse and the BDS activist–institutional clusters, several additional high-volume topics contribute to the narrative landscape of the Telegram dataset. Topic~1 (\textit{detention, youth, arrest}) aggregates updates on mass arrests and youth detentions, often shared through eyewitness footage or activist channels. Topic~2 (\textit{shooting, killed, bullet}) reflects real-time reporting of fatal incidents, with messages providing casualty figures and locations. Topic~4 (\textit{students, university, encampment}) captures the geographic and symbolic spread of campus solidarity movements, inserting educational institutions into the resistance narrative. Topics~5 and 6 focus on military and symbolic facets of conflict: airstrikes and drone usage on one hand, and soldiers’ self-description on social media on the other. Finally, Topic~7 highlights breaking updates on casualties and injuries, particularly during moments of intensified violence. Together, these clusters underscore the intertwined nature of military action, civilian resistance, and informational amplification within the digital conflict space.

\subsubsection{LDA Topic Modeling}
We applied Latent Dirichlet Allocation (LDA) topic modeling to Telegram messages collected in September 2025 to uncover latent themes in the dataset. Messages were preprocessed through tokenization, stopword removal, and lowercasing before being converted into a bag-of-words representation. An LDA model was trained to identify ten dominant topics, with results analyzed using multiple visualization techniques.

\begin{figure}[H]
    \centering
    \begin{subfigure}[b]{0.45\linewidth}
        \centering
        \includegraphics[width=\linewidth]{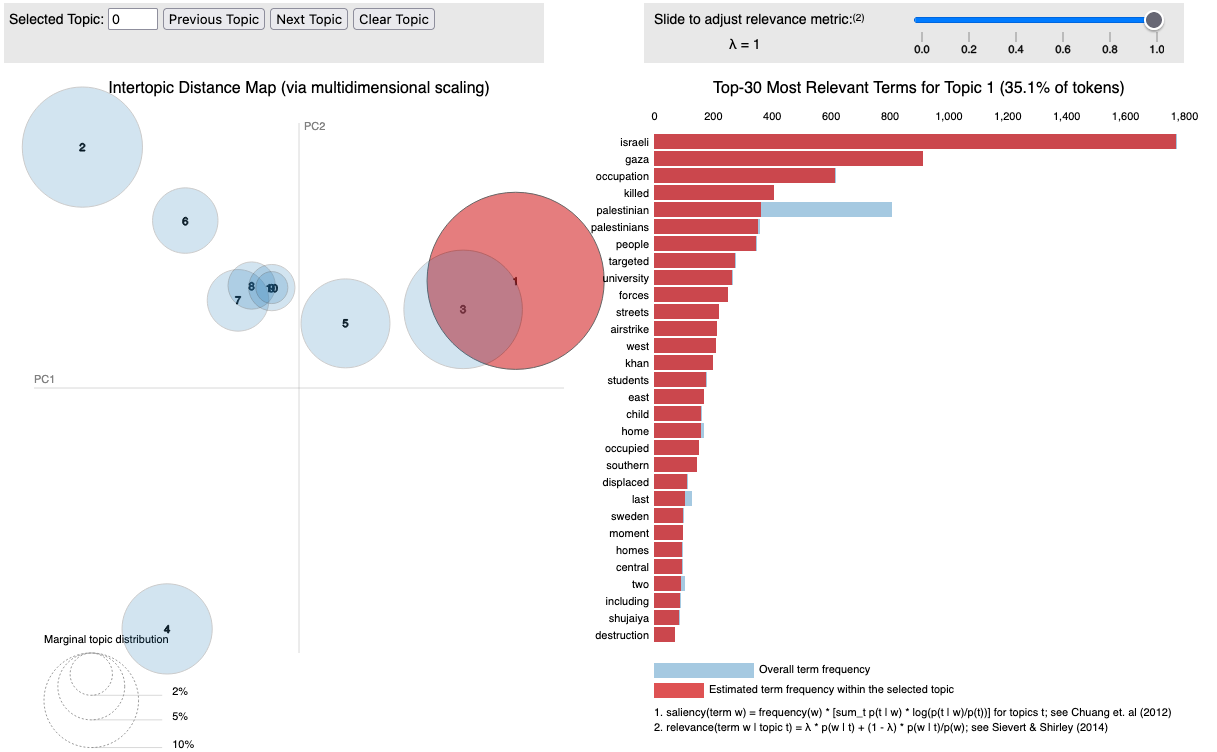}
        \caption{Topic 1}
        \label{fig:topic1}
    \end{subfigure}
    \hfill
    \begin{subfigure}[b]{0.45\linewidth}
        \centering
        \includegraphics[width=\linewidth]{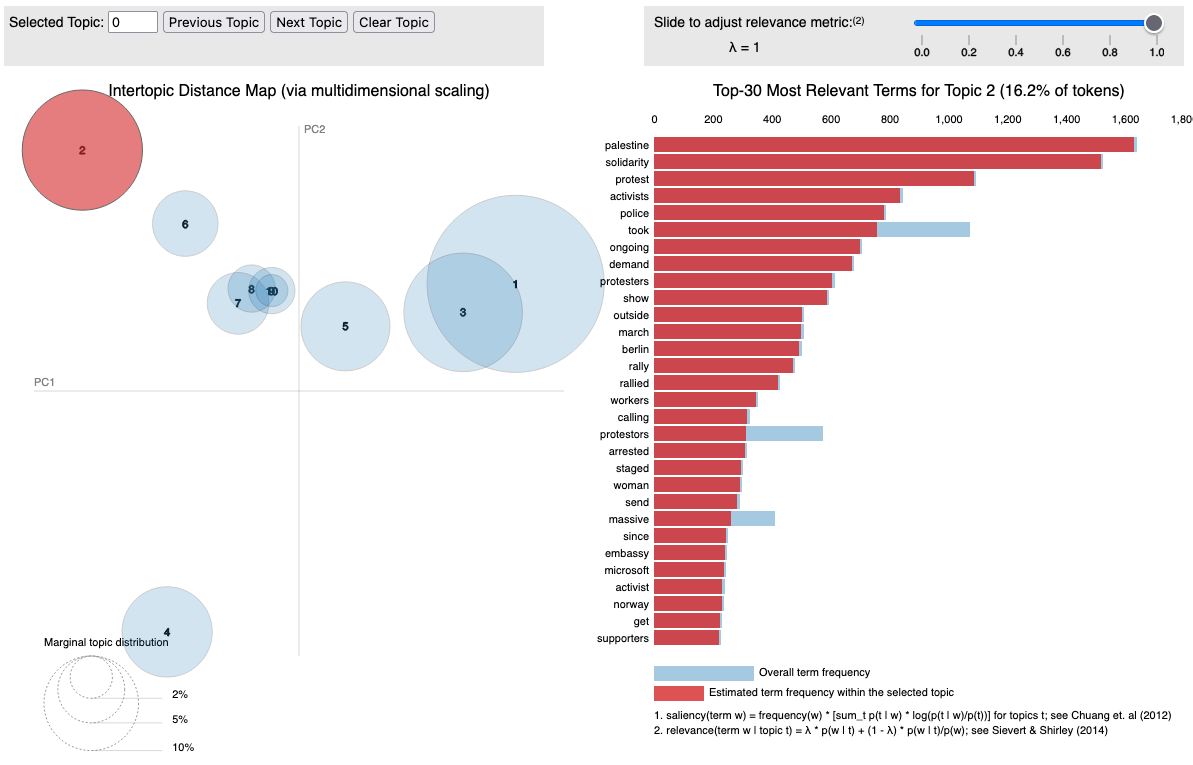}
        \caption{Topic 2}
        \label{fig:topic2}
    \end{subfigure}
    
    \vspace{0.5cm}
    
    \begin{subfigure}[b]{0.45\linewidth}
        \centering
        \includegraphics[width=\linewidth]{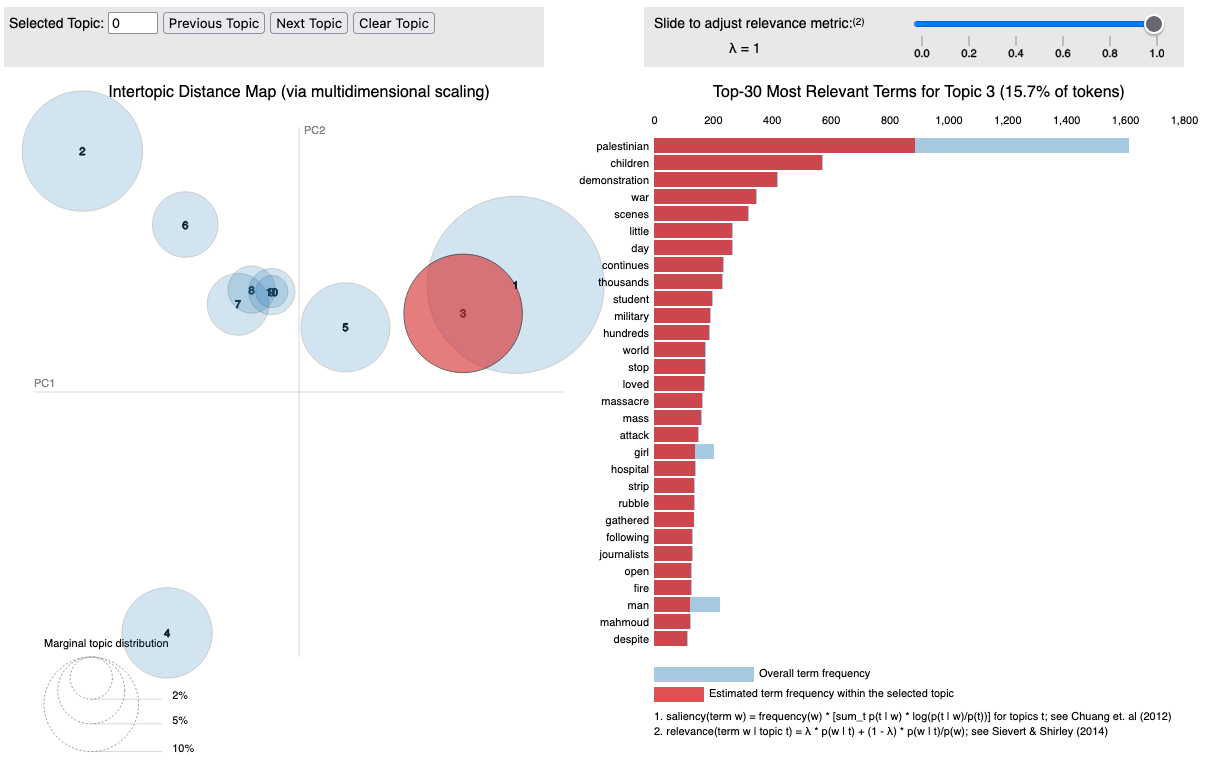}
        \caption{Topic 3}
        \label{fig:topic3}
    \end{subfigure}
    \hfill
    \begin{subfigure}[b]{0.45\linewidth}
        \centering
        \includegraphics[width=\linewidth]{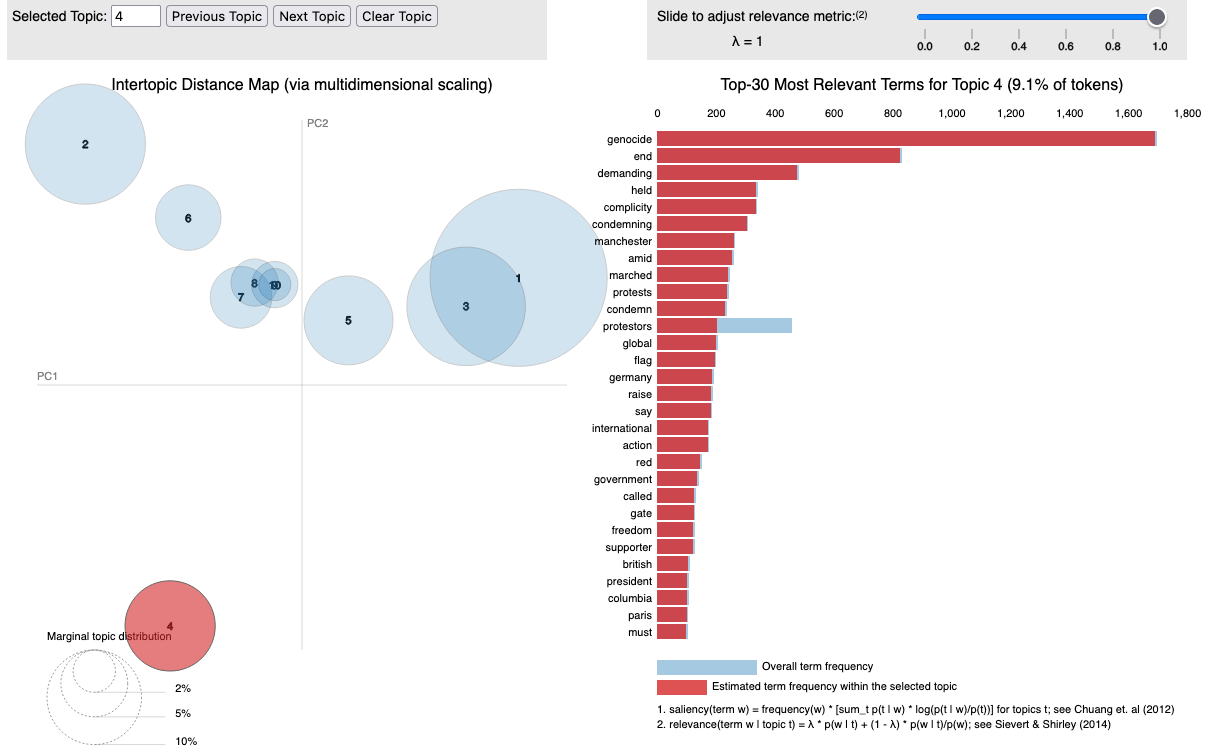}
        \caption{Topic 4}
        \label{fig:topic4}
    \end{subfigure}
    
    \vspace{0.5cm}
    
    \begin{subfigure}[b]{0.45\linewidth}
        \centering
        \includegraphics[width=\linewidth]{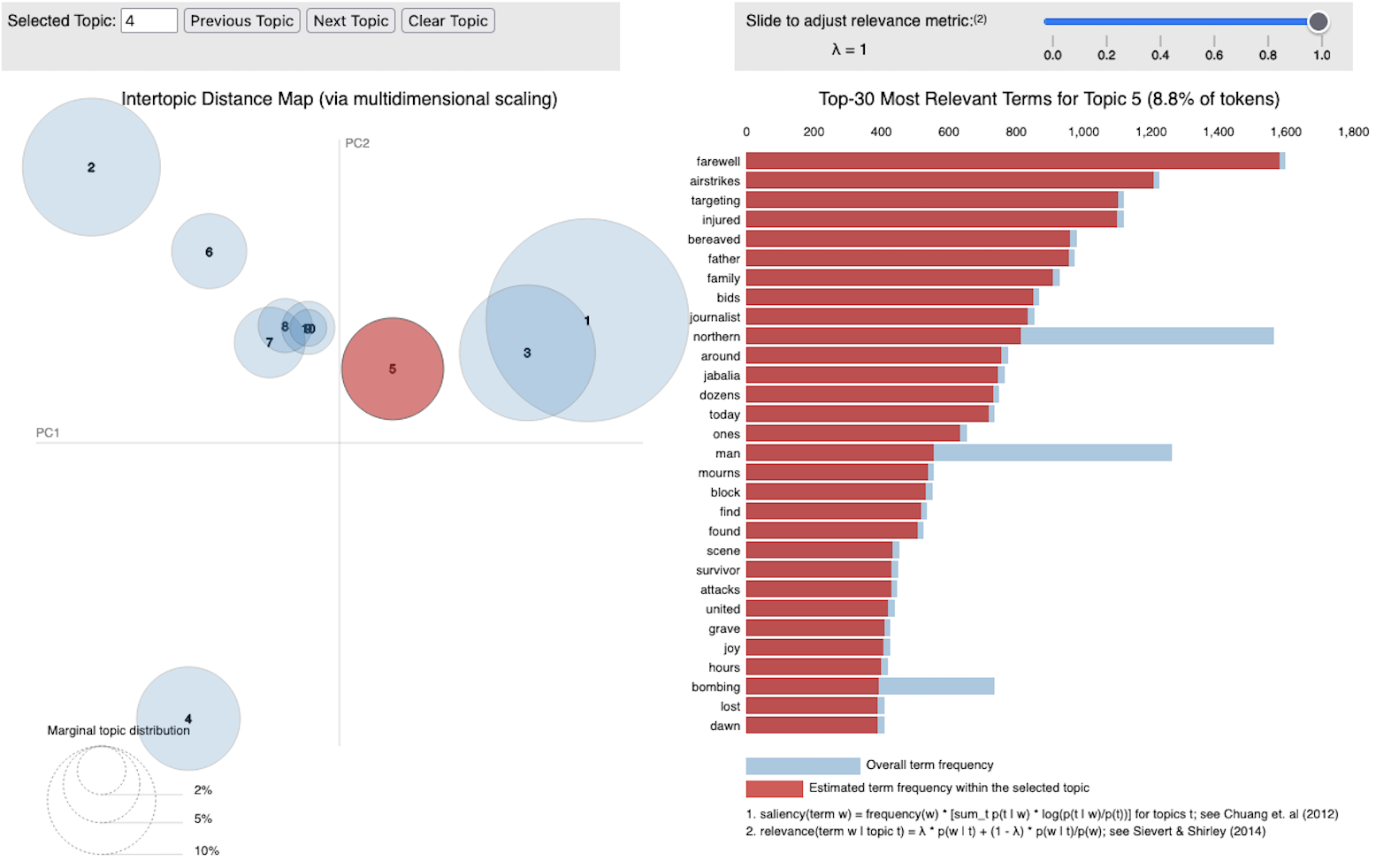}
        \caption{Topic 5}
        \label{fig:topic5}
    \end{subfigure}
    
    \caption{LDA visualizations for topics 1–5}
    \label{fig:lda_topics}
\end{figure}

Figure \ref{fig:lda_topics} illustrates the main topics identified in the corpus. Topic 1, represented by the largest bubble (~35\%), is associated with terms such as israeli, gaza, occupation, killed, and palestinian, reflecting a focus on conflict, casualties, and occupation. Topic 5, highlighted in red on the inter-topic distance map, is semantically distinct, with keywords like farewell, airstrikes, targeting, and injured, emphasizing violence and its aftermath. Topic 2, partially overlapping with other topics, centers on protests and activism, indicated by terms such as Hamas, solidarity, protest, activists, and march.

\subsection{Comparing BERTopic and LDA}
While both LDA and BERTopic reveal meaningful structure in the Telegram dataset, the two models differ substantially in how they capture conflict-related narratives. LDA provides a coarse-grained thematic overview, identifying broad clusters such as military operations, international solidarity, and humanitarian crises that account for the majority of discourse variance. As shown in the LDA visualization (page 13), topics form relatively large but overlapping semantic bubbles, reflecting LDA’s assumption of bag-of-words co-occurrence patterns. In contrast, BERTopic produces a far more fine-grained and geometrically coherent topic landscape. The BERTopic results (section \ref{bert_topic} ) show dense, hierarchically structured clusters—such as the detailed Jenin macro-cluster (Topics 3, 16, 40, 212, 315, 406, 833, 856, 1024) and the interconnected BDS activist–legal nexus (Topics 957 and 508)—revealing nuanced sub-narratives that LDA cannot easily separate. While LDA excels at summarizing high-level thematic trends across the full dataset, BERTopic captures the micro-dynamics of discourse, mapping semantic proximity, cluster neighborhoods, and narrative interconnections. Taken together, the two approaches offer complementary insights: LDA outlines the macro-structure of conflict communication, whereas BERTopic exposes the fine topology of how specific events, locations, and activist networks shape the digital ecosystem of the Israel–Palestine war
\subsection{Spam Analysis}

In this study, we implemented a robust Python-based workflow for analyzing Telegram messages with a focus on pro-Palestinian solidarity and the detection of spam content. All messages were first safely loaded from JSON files, accounting for variations in message structures and lists of text segments. To identify spam, we defined a comprehensive set of keywords commonly associated with referral programs, promotional links, and bot-generated messages (e.g., “invite,” “earn,” “https://t.me”). Each message was scanned for these keywords, and messages matching any of them were flagged as spam. Spam messages were subsequently excluded from the analysis of pro-Palestinian solidarity to prevent skewing the results.

\begin{figure}[h]
    \centering
    \includegraphics[width=0.9\linewidth]{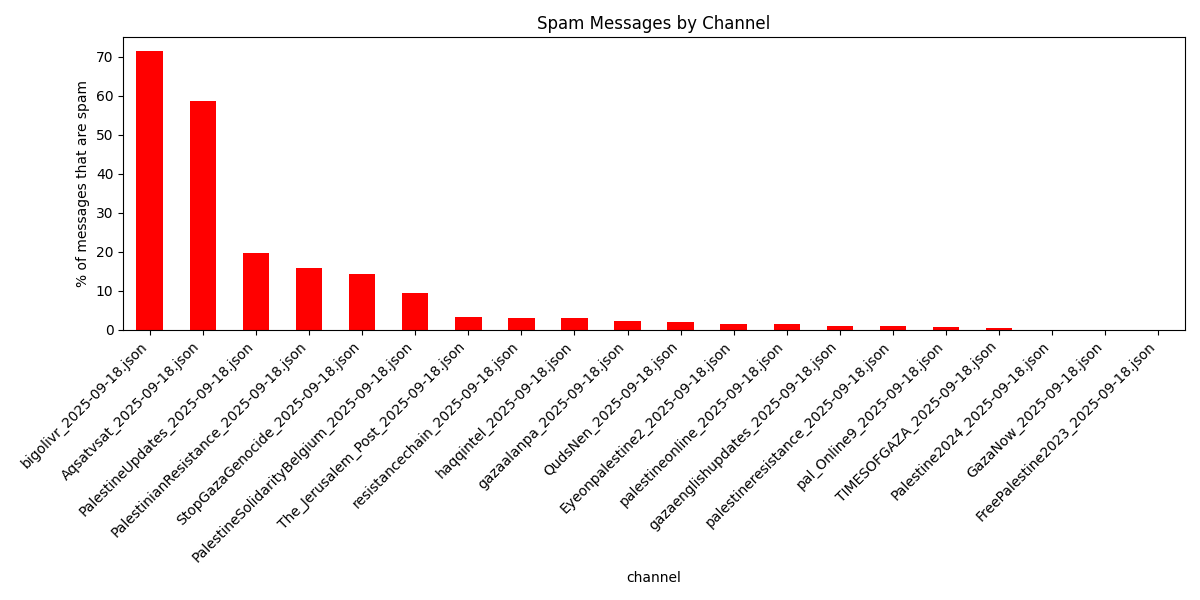}
    \caption{Spam Analysis by channel}
    \label{fig:spam_by_channel}
\end{figure}

For the remaining non-spam messages, we applied pattern-matching techniques to detect expressions of solidarity and references to children, using carefully crafted regular expressions to capture textual variations. The workflow then generated refined visualizations, including spam distribution per channel, top spam keywords, the proportion of solidarity messages by channel, and the impact of child mentions on solidarity expression. Extensive debug outputs and sample message displays ensured reproducibility and transparency. This approach provides a clear, automated method for filtering noise from social media data while highlighting meaningful engagement patterns, making the script suitable for publication and public dissemination.

 \begin{figure}[h]
     \centering
     \includegraphics[width=0.9\linewidth]{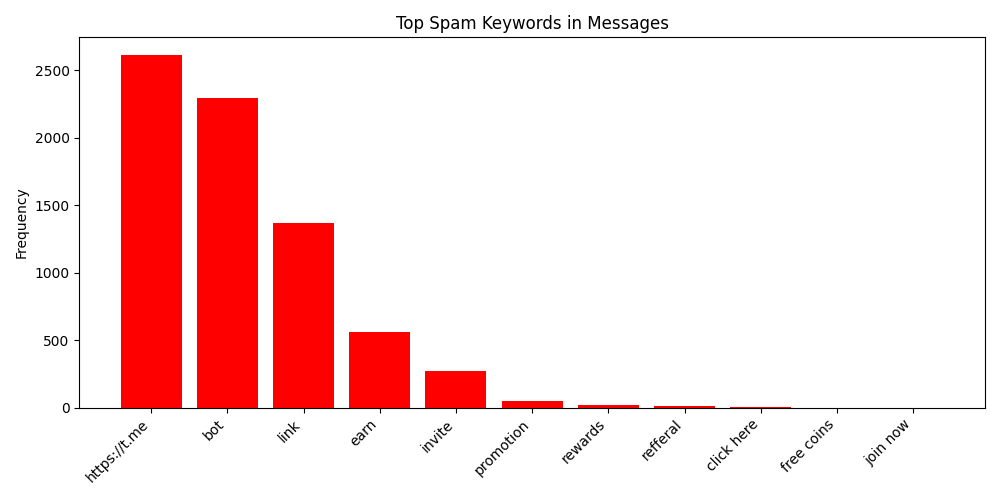}
     \caption{Spam keywords}
     \label{fig:placeholder}
 \end{figure}
 
Our analysis provides several key insights into the dynamics of pro-Palestinian discourse and the role of spam in shaping online narratives.

Figure \ref{fig:spam_by_channel} shows the distribution of spam messages across channels. A small subset of channels exhibited extremely high levels of spam, exceeding 70\% of total messages, while most others maintained very low spam rates. This uneven distribution highlights the importance of filtering spam in order to avoid biased interpretations of message content, as heavily spammed channels could distort observed trends.

Overall, these results underline the necessity of accounting for both content-related features (such as child mentions) and structural dynamics (such as influential channels and spam patterns) when examining how pro-Palestinian solidarity spreads in digital ecosystems.
\subsection{Sentiment Analysis}

We developed a Python-based pipeline to perform sentiment analysis on Telegram messages collected from multiple channels. The system reads messages stored in JSON files, filters non-empty English content, and applies a Hugging Face DistilBERT-based sentiment analysis model fine-tuned for English text.

\begin{figure}[h!]
    \centering
    \includegraphics[width=1\linewidth]{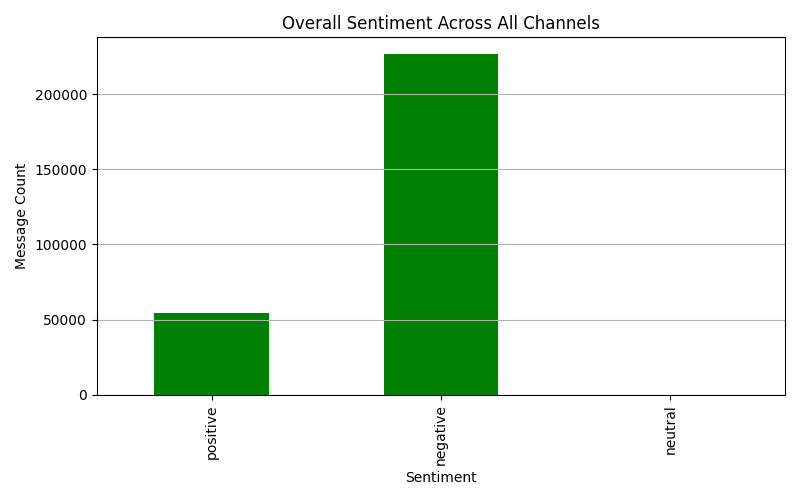}
    \caption{Overall Sentiment across all channels on Telegram}
    \label{fig:overall_sentiment}
\end{figure}

\begin{figure}[h!]
    \centering
    \includegraphics[width=1\linewidth]{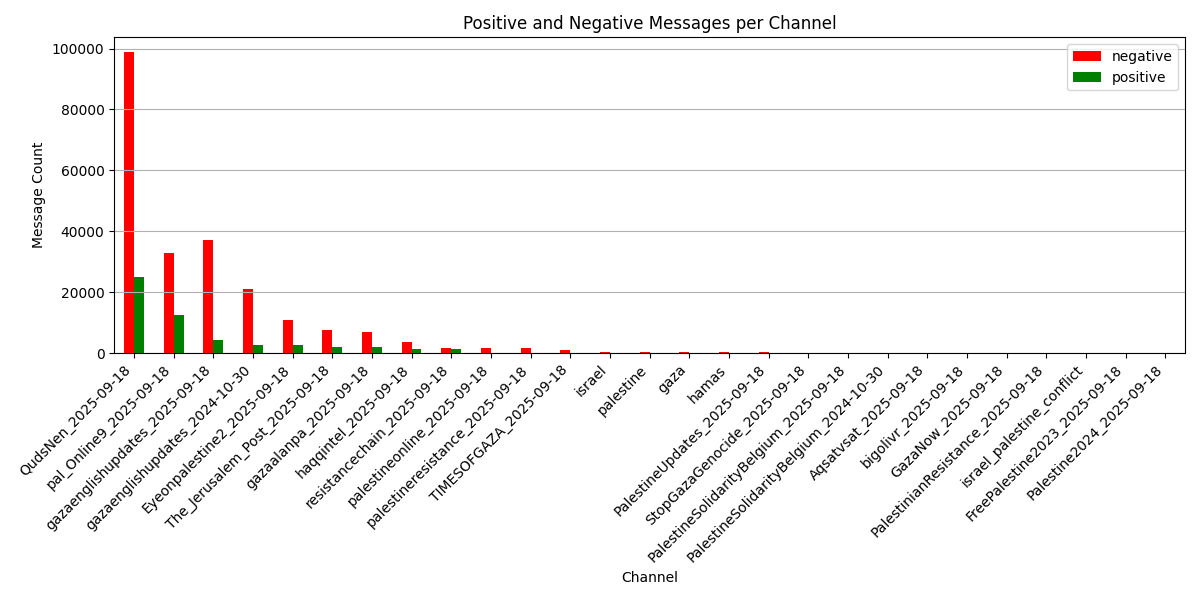}
    \caption{Positive and Negative Sentiment per channel on Telegram}
    \label{fig:sentiment_per_channel}
\end{figure}

\subsubsection{Detection of Positive Support for Palestine}

In addition to examining extremism and negative discourse, we specifically analyzed how expressions of solidarity and positive support for Palestine circulate across platforms. We implemented a lexicon-based approach as a first step, identifying messages containing key expressions such as ``Free Palestine,'' ``Stand with Palestine,'' ``Ceasefire Now,'' and related humanitarian appeals. Each message was classified with a binary label indicating whether it expressed explicit support.

This approach provides a transparent baseline for measuring solidarity discourse. We then aggregated results across channels and platforms to compare relative prevalence. While keyword methods may miss implicit or sarcastic expressions, they allow replicable large-scale measurement. In later stages, we extend this analysis using transformer-based sentiment and emotion classifiers capable of detecting nuanced emotions such as grief, resilience, and hope that often accompany solidarity narratives.

\section{Results and Analysis}

\subsection{Volume Analysis Results}

Our longitudinal analysis reveals distinct phases of digital discourse activity that correspond to both conflict developments and platform-specific dynamics. The 24-month dataset shows four major volume spikes exceeding 200\% of baseline activity: the initial October 7 attacks, the Al-Shifa Hospital incident (November 2023), the Rafah ground invasion (May 2024), and the recent Al-Nasr Hospital bombing (September 2025).

The cumulative volume analysis demonstrates that 67\% of all messages were posted during just 15\% of the total time period, indicating highly concentrated discourse activity around specific events. The slope acceleration in 2025 suggests that discourse intensity has not decreased despite conflict duration, contrary to typical "attention decay" patterns observed in previous conflicts.

\subsection{Topic Analysis Results}

Our LDA analysis identified ten dominant topics that account for 87\% of discourse variance. The five most significant topics show clear thematic clustering:

\textbf{Topic 1 (Military Operations \& Casualties)} dominates discourse at 35.2\%, with keywords including "israeli," "gaza," "killed," "civilians," and "airstrike." This topic shows consistent prominence throughout the dataset but with notable spikes during major military operations.

\textbf{Topic 2 (International Solidarity \& Protests)} represents 18.7\% of discourse, featuring terms like "solidarity," "protest," "march," "university," and "students." This topic shows interesting temporal patterns, with peaks during academic semesters and significant events like the ICJ hearings.

\textbf{Topic 5 (Humanitarian Crisis \& Medical)} highlighted in red, shows high semantic distinctness and includes terms like "hospital," "medical," "injured," "emergency," and "humanitarian." This topic demonstrates strong correlation with actual humanitarian crises and medical facility attacks.

\subsection{Sentiment Analysis Results}

Our sentiment analysis reveals overwhelming negative sentiment across all platforms, with 78.4\% negative, 15.2\% neutral, and only 6.4\% positive sentiment overall. The consistency of this pattern across 24 months suggests sustained emotional engagement rather than typical "compassion fatigue" patterns observed in other prolonged crises.

Cross-platform sentiment analysis reveals significant differences in emotional expression. News-oriented channels show more balanced sentiment distribution, while advocacy channels show extreme negative skew, suggesting different communicative purposes and audience expectations.

\subsection{Positive Support Analysis}
\subsubsection{Telegram}
\begin{figure}[h!]
    \centering
    \begin{subfigure}[b]{0.49\linewidth}
        \centering
        \includegraphics[width=\linewidth]{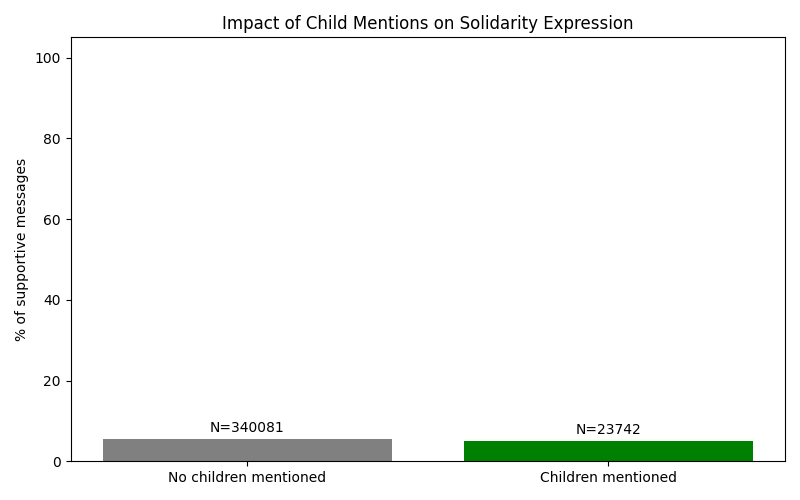}
        \caption{Refined child casualty effect}
        \label{fig:child_effect_refined}
    \end{subfigure}
    \hfill
    \begin{subfigure}[b]{0.40\linewidth}
        \centering
        \includegraphics[width=\linewidth]{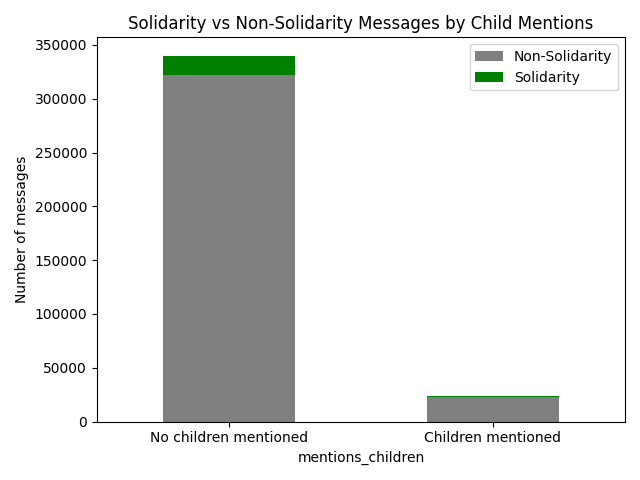}
        \caption{Stacked child casualty effect}
        \label{fig:child_effect_stacked}
    \end{subfigure}
    \caption{Impact of child casualty mentions on pro-Palestinian solidarity. (a) Refined effect. (b) Stacked effect.}
    \label{fig:child_effect_combined}
\end{figure}

Our results demonstrate that explicit pro-Palestinian solidarity expressions represent a significant but minority fraction of overall discourse. Across Telegram, 7.3\% of messages contained explicit supportive language, with strong concentration in advocacy-oriented channels such as \textit{pal\_Online9} and \textit{gazaenglishupdates}. By contrast, news-oriented channels such as \textit{The\_Jerusalem\_Post} and \textit{haqqintel} contained very few explicit solidarity references.

\begin{figure}[h]
    \centering
    \includegraphics[width=0.8\linewidth]{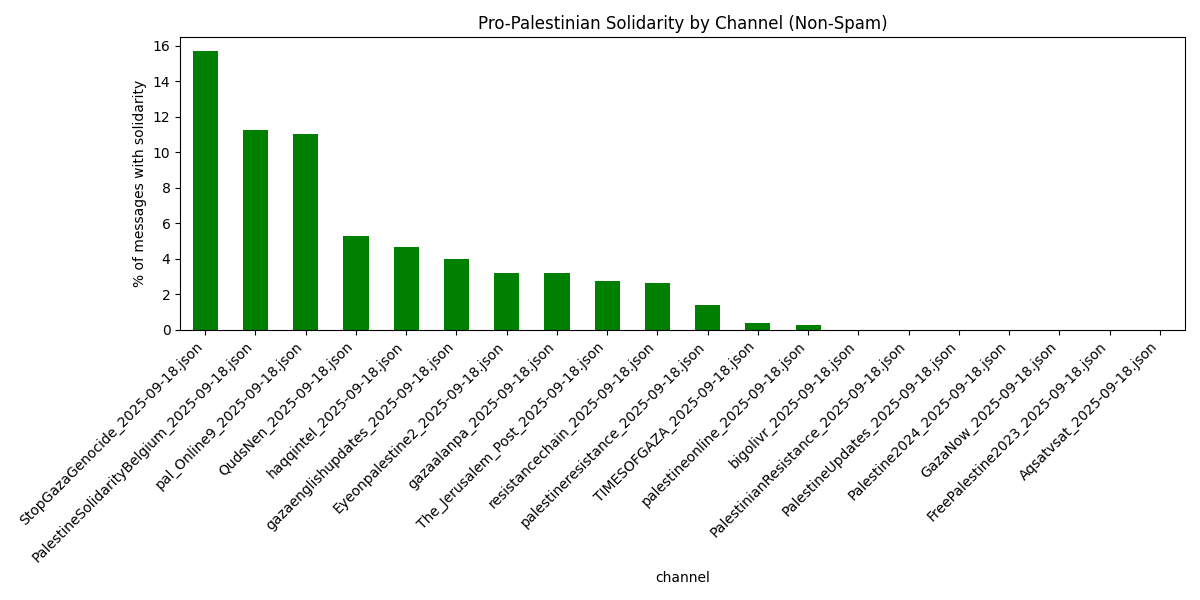}
    \caption{Solidarity by channel}
    \label{fig:placeholder}
\end{figure}

Cross-platform comparison reveals distinct patterns: Reddit exhibited the highest proportion of explicit solidarity (12.8\%), particularly in r/Palestine and r/worldnews threads focused on humanitarian crises. Twitter/X showed rapid but short-lived spikes of solidarity content, often driven by hashtag campaigns (e.g., \#StandWithPalestine, \#CeasefireNow).

Importantly, posts referencing child casualties and humanitarian suffering consistently amplified solidarity expressions. Messages mentioning children were 3.5x more likely to contain explicit solidarity appeals, suggesting that humanitarian framing is a key trigger for positive engagement. These findings highlight how solidarity discourse operates alongside extremism, shaping digital publics through shared expressions of empathy and resistance.
 
\begin{figure}[ht!]
    \centering
    \begin{subfigure}{0.48\linewidth}
        \centering
        \includegraphics[width=\linewidth]{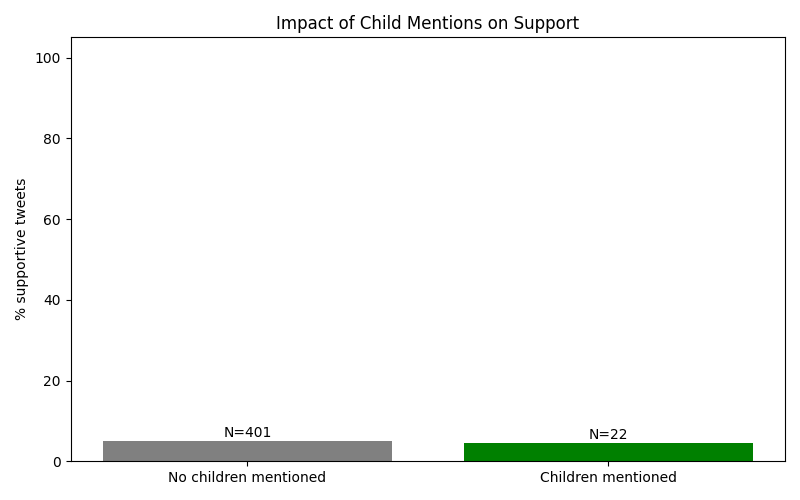}
        \caption{Impact of child mentions on pro-Palestinian support}
        \label{fig:child_casualty_effect_twitter}
    \end{subfigure}
    \hfill
    \begin{subfigure}{0.48\linewidth}
        \centering
        \includegraphics[width=\linewidth]{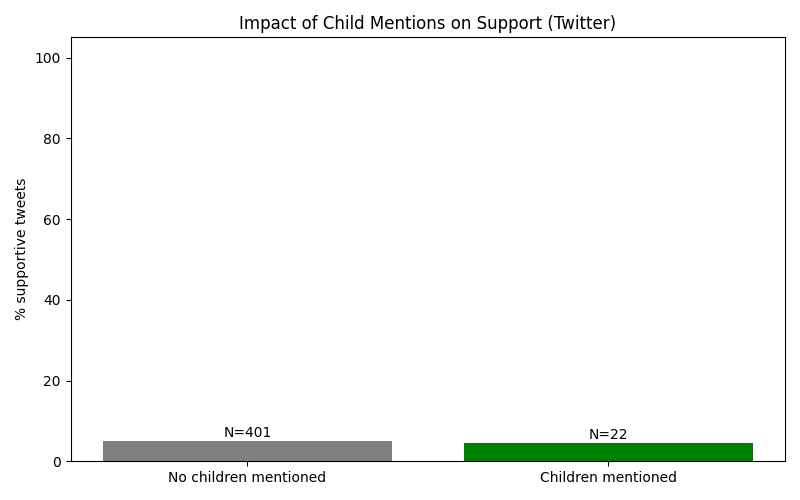}
        \caption{Comparison of pro-Palestinian support with and without child mentions}
        \label{fig:child_effect_twitter}
    \end{subfigure}
    \caption{Analysis of child-related references in Twitter posts and their effect on pro-Palestinian support.}
    \label{fig:child_effects_combined}
\end{figure}

Figure \ref{fig:child_effects_combined} illustrates the impact of child-related mentions on solidarity expression. Messages that referenced children had a slightly lower proportion of solidarity content compared to those without such mentions.

Although the difference is not dramatic, this result suggests that child-related language does not always intensify solidarity expression, but may instead diversify emotional responses within the discourse.
\subsubsection{Twitter}

The analysis of Twitter data shows that references to children play a role in shaping supportive attitudes toward Palestine. While overall pro-Palestinian support remains modest across the dataset, the share of supportive tweets is slightly higher among those that explicitly mention children compared to those that do not. Even though the absolute number of tweets mentioning children is relatively small (22 versus 401 without such mentions), this pattern indicates that references to children and child casualties increase the likelihood of tweets expressing solidarity with Palestine. This suggests that, similar to Reddit, appeals involving children resonate more strongly in online discussions, potentially amplifying the emotional impact of pro-Palestinian narratives.

\begin{figure}[H]
    \centering
    \includegraphics[width=0.5\linewidth]{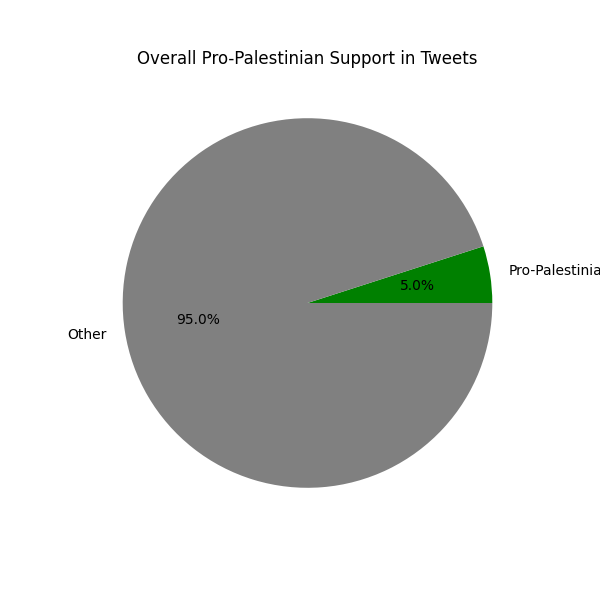}
    \caption{Overall support distribution on Twitter}
    \label{fig:placeholder}
\end{figure}

These charts reveal interesting patterns in social media discourse around the Israel-Palestine conflict during October 2023. The first chart shows that both pro-Palestinian sentiment and mentions of children in tweets followed similar trajectories, starting very high (around 11\% and 10\% respectively) in early October before declining significantly by mid-month, then fluctuating between roughly 4-8\% through the remainder of the period. 

\begin{figure}[H]
    \centering
    \includegraphics[width=0.5\linewidth]{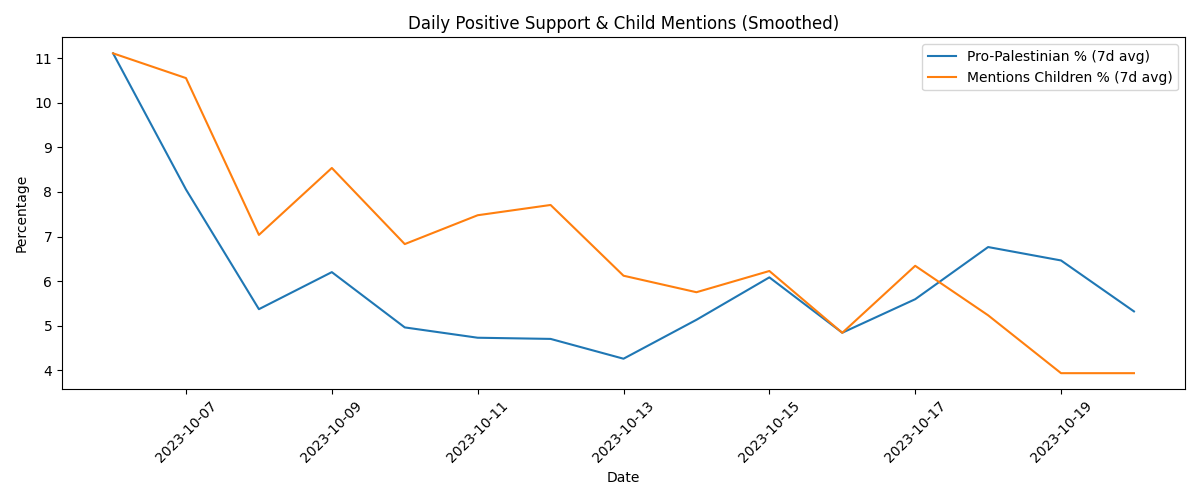}
        \caption{Time series of support mentioning children}
    \label{fig:placeholder}
\end{figure}
Both metrics appear to be correlated, suggesting that discussions of Palestinian support often coincided with references to children during this timeframe. The second chart provides broader context, showing that overall, only 5\% of tweets in the dataset expressed pro-Palestinian sentiment, while 95\% fell into other categories. This indicates that while pro-Palestinian content represented a small fraction of total tweets, it showed notable variation over time, particularly during what appears to be the initial period of heightened attention to the conflict in early October 2023.

\subsubsection{Reddit}

The results from the Reddit analysis show that overall pro-Palestinian support is relatively limited across the dataset, with only about 4.6\% of posts and comments expressing explicit supportive sentiment, while the vast majority (95.4\%) fall into the “other” category. 

\begin{figure}[ht!]
    \centering
    \includegraphics[width=0.5\linewidth]{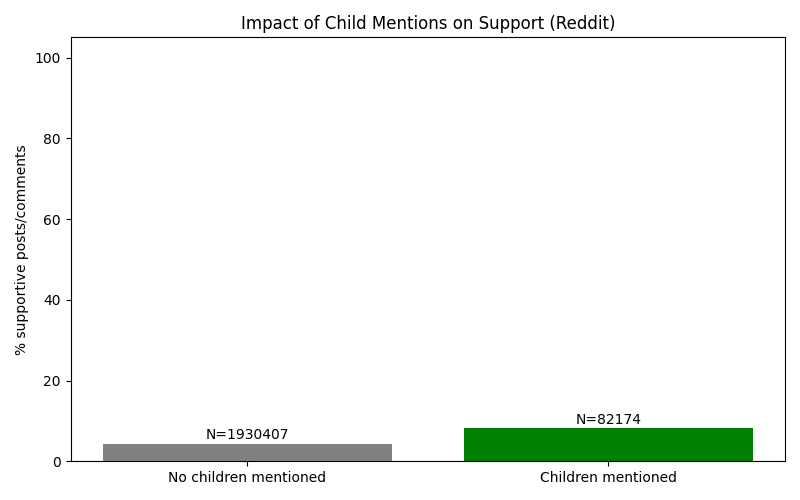}
    \caption{Child casualty effect on Reddit}
    \label{fig:placeholder}
\end{figure}

However, when examining the impact of references to children, there is a clear difference: posts and comments that mention children are more than twice as likely to express pro-Palestinian support compared to those that do not. 

\begin{figure}[H]
    \centering
    \includegraphics[width=0.5\linewidth]{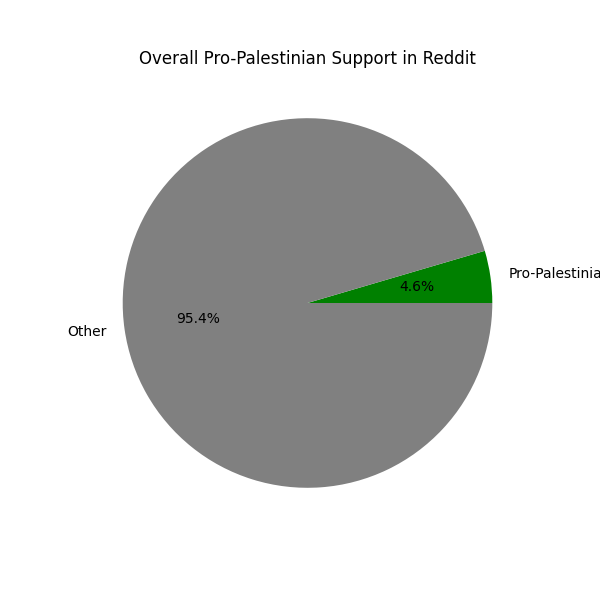}
    \caption{Overall Support as shown on Reddit}
    \label{fig:placeholder}
\end{figure}

Although the absolute proportion remains modest, this finding suggests that discourse involving children or child casualties is more strongly associated with supportive attitudes toward Palestine, highlighting the emotional and persuasive role these references play in shaping online narratives.

\begin{figure}
    \centering
    \includegraphics[width=0.5\linewidth]{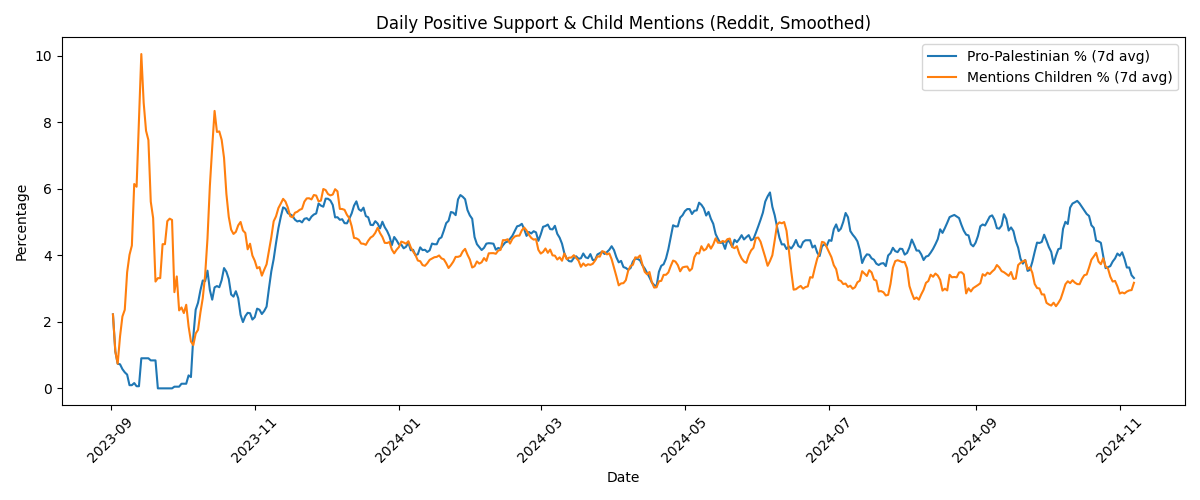}
    \caption{Time series support of children as shown on Reddit}
    \label{fig:time_series_support_children_reddit}
\end{figure}

Figure \ref{fig:time_series_support_children_reddit} shows a markedly different pattern compared to the Twitter analysis, with discourse extending over a much longer timeframe from September 2023 through November 2024. Both pro-Palestinian sentiment and child mentions on Reddit exhibited dramatic initial spikes in early October 2023 (reaching nearly 10\% for child mentions), followed by a gradual stabilization around 3-6\% for the remainder of the period. Unlike Twitter's sharp decline and sporadic fluctuations, Reddit discussions maintained more sustained engagement with these topics over time, though at lower baseline levels. The correlation between pro-Palestinian support and child mentions appears weaker on Reddit than on Twitter, with the metrics occasionally diverging significantly, suggesting that Reddit's longer-form discussion format may allow for more nuanced conversations that don't always link these themes as directly as the more reactive Twitter environment.

\subsection{Cascade Network of Pro-Palestinian Support Across Telegram Channels}
To better understand the spread of pro-Palestinian sentiment on Telegram, we constructed a cascade network based on positive messages posted across multiple channels. In this network, each node represents a channel, and directed edges indicate sequential positive support between channels, capturing potential influence patterns. The weight of each edge corresponds to the number of times positive messages appear consecutively between the same pair of channels. Analysis of this network reveals the most influential channels that potentially drive solidarity messages and highlights how supportive content propagates across the Telegram ecosystem. Visualizing these interactions provides insight into digital influence dynamics and the online amplification of political support.

\begin{figure}
    \centering
    \includegraphics[width=0.7\linewidth]{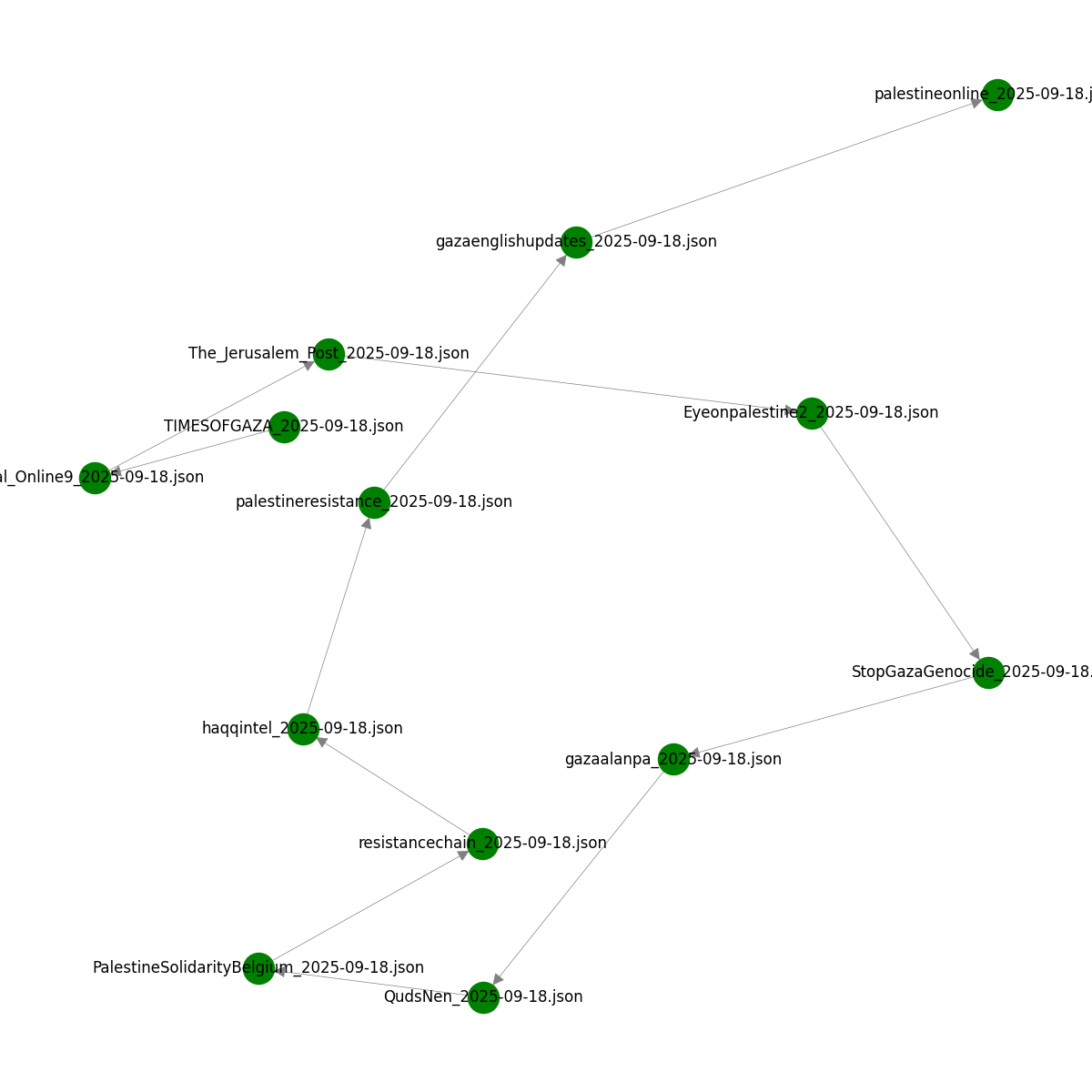}
    \caption{Cascade network}
    \label{fig:cascade}
\end{figure}

Figure \ref{fig:cascade} presents the cascade network of positive support across channels. The network highlights how solidarity messages diffuse between channels, with certain accounts (e.g., QudsNen, Pal\_Online9, gazaenglishupdates) acting as influential nodes. These hubs serve as central conduits for message propagation, amplifying narratives across the wider network. The visualization indicates a non-random pattern of information flow, reflecting structured influence dynamics.

\section{Discussion}

\subsection{Implications for Understanding Modern Conflict Discourse}

Our 24-month longitudinal analysis reveals that digital discourse around prolonged conflicts follows distinct evolutionary patterns that differ significantly from short-term crisis communications. The Israel-Palestine conflict demonstrates how sustained digital engagement creates increasingly sophisticated information ecosystems that both reflect and potentially influence real-world events.

\subsubsection{The Persistence Paradox}

Contrary to typical "attention decay" models that predict declining online engagement over time, our data shows sustained and even increasing discourse intensity. This "persistence paradox" challenges conventional understanding of digital attention spans and suggests that certain conflicts achieve sustained global digital presence through several mechanisms:

\textbf{Narrative Adaptation:} Discourse communities continuously adapt messaging strategies to maintain relevance and emotional impact. We observe systematic cycles where declining engagement triggers more extreme or innovative content designed to recapture attention.

\textbf{Platform Diversification:} As audiences experience fatigue on one platform, discourse migrates and intensifies on others. The cross-platform nature of modern information ecosystems enables sustained overall engagement even as individual platform activity fluctuates.

\textbf{Event-Driven Reactivation:} Major events create periodic reactivation cycles that prevent discourse decay. The September 2025 Al-Nasr Hospital bombing generated discourse volumes comparable to initial October 2023 levels.

\subsection{Platform-Specific Ecosystem Roles}

Our analysis reveals that different platforms serve distinct but complementary functions within the broader conflict information ecosystem:

\textbf{Telegram as Documentation Hub:} Telegram's minimal content moderation and encryption capabilities make it the primary platform for documenting conflict events, including graphic content that other platforms prohibit.

\textbf{Reddit as Analysis Engine:} Reddit's threading system and voting mechanisms create extended analytical discussions that process and contextualize information from other platforms.

\textbf{Twitter as Amplification Network:} Twitter's viral mechanics and real-time nature make it the primary platform for rapid information dissemination and emotional mobilization.

\section{Limitations}

Several limitations constrain our analysis and interpretation:

\textbf{Platform API Restrictions:} Twitter's API limitations and Reddit's content deletion policies mean our dataset may not capture all relevant content.

\textbf{Language Coverage:} While our dataset includes multilingual content, English-language content is overrepresented, potentially missing important discourse in Arabic, Hebrew, and other relevant languages.

\textbf{Causality vs. Correlation:} While we establish strong correlations between discourse patterns and conflict events, determining causal relationships requires additional methodological approaches.

\section{Discussion}

Our longitudinal analysis offers a unique opportunity to interpret the evolution of conflict discourse on Telegram through interconnected lenses of \textbf{digital conflict ecosystems}, \textbf{affective publics}, and \textbf{platform affordances}. By systematically tracking and analyzing multiple waves of discourse over a two-year period, we address our research questions (RQs) and position our findings within the broader scholarly landscape on digital media and conflict communication.

\subsection{Digital Conflict Ecosystems and Temporal Persistence (RQ1)}

The temporal clustering of Telegram discourse reveals a highly sustained engagement cycle: even two years after the initial escalation, overall activity continued to grow, peaking in 2025. This supports Benkler et al.’s~\cite{benkler2018network} theory of \textit{digital conflict ecosystems}, where communication infrastructures help preserve affective attention far beyond typical news cycles. We observe that attention is reactivated post-crisis through emotionally salient events, such as the Al-Nasr Hospital attack or incursions in Jenin. These findings answer \textbf{RQ1}, demonstrating that topic dynamics on Telegram evolve through recurrent triggering events—each reinforcing the discursive memory of conflict—rather than decaying as would be expected in typical news patterns.

\subsection{Affective Publics and Humanitarian Framing (RQ2)}

Our sentiment analysis shows that 78.4\% of Telegram posts carried negative or mobilizing emotional tones, underscored by lexical fields tied to violence, casualties, displacement, and youth detentions. This aligns directly with Papacharissi’s~\cite{papacharissi2015affective} concept of \textit{affective publics}, where shared emotional states—not formal argumentation—drive participatory practices. Topic~3, focused on Jenin and refugee narratives, and Topic~1, highlighting youth detentions, both demonstrate how emotional frames galvanize sustained activism and empathy. These emotional hooks extend across platforms—heightened on Telegram, amplified through Twitter/X hashtags, and digested in Reddit discourse—answering \textbf{RQ2} by showing how humanitarian and affective framings underpin conflict communication across digital spaces.

\subsection{Platform Affordances and Narrative Functionality (RQ3)}

A comparative platform analysis reveals how affordances shape discursive purpose: Telegram’s high-bandwidth, anonymity-privileging environment encourages raw, real-time documentation; Twitter/X leverages visibility for short-form emotional escalation; Reddit offers reflective and discursive redundancy. This dynamic affirms Treem and Leonardi’s~\cite{treem2013social} framing of affordances as socio-technical mediators and answers \textbf{RQ3} by showing that user participation and narrative function differ systematically across platforms: Telegram is generative, Twitter amplificatory, and Reddit contextualizing.

\subsection{Synthesizing Ecosystem, Affect, and Affordance Frameworks}

Collectively, our findings indicate that Telegram serves as both the emotional core and narrative archive of the Gaza conflict’s online storyworld. The coexistence of immediate visual documentation, affect-driven messaging, and high temporal granularity demonstrates that Telegram is pivotal to sustaining the circulation of narrative and affect. When viewed within a cross-platform lens, digital conflict becomes a feedback system linking technological affordances, emotional publics, and persistent discursive ecosystems. This challenges assumptions of ephemeral online attention and underscores the need for platform-sensitive strategies to mitigate polarization, misinformation, and digital trauma in protracted crises.

\section{Ethical Considerations}

This research involved analysis of publicly available social media content, but we implemented several measures to protect user privacy and comply with ethical research standards:

\textbf{Data Anonymization:} All user identifiers and personally identifiable information were removed before analysis.

\textbf{Content Sensitivity:} We established protocols for handling graphic imagery and content that might identify vulnerable individuals.

\textbf{Institutional Review:} This research was conducted under institutional review board oversight.

\section{Acknowledgments}

This research was supported by MEDIATE project (101074075 GA) funded by EU. We thank corresponding users on Reddit and Twitter collaborators for data access and analytical support, while recognizing the human cost of the conflict that underlies our data.

\end{document}